\documentclass[onecolumn,english,superscriptaddress,prl,nofootinbib,nobibnotes]{revtex4-1}
\usepackage{lmodern}
\usepackage[T1]{fontenc}
\usepackage[utf8]{inputenc}

\usepackage{color}
\usepackage{babel}
\usepackage{amsmath}
\usepackage{amssymb}
\usepackage{bm}
\usepackage{graphicx}
\usepackage{empheq}

\usepackage[unicode=true,pdfusetitle,
bookmarks=true,bookmarksnumbered=false,bookmarksopen=false,
breaklinks=false,hypertexnames=false,pdfborder={0 0 1},backref=false,colorlinks=true]
{hyperref}
\hypersetup{
	pdfborderstyle=,urlcolor=orange,linkcolor=blue,citecolor=red}

\makeatletter

\usepackage{etoolbox}

\usepackage{bigints}

\usepackage{algorithm}
\usepackage{algpseudocode}

\usepackage{color}
\usepackage{babel}
\usepackage{units}

\usepackage[normalem]{ulem}
\usepackage{bigints}

\usepackage{comment}

\usepackage{tikz}
\usepackage{color}
\usepackage{dsfont}
\usepackage[cal=boondoxo]{mathalfa}
\usepackage{grffile}
\usepackage[caption=false]{subfig}

\catcode`,\active

\catcode`\,12

\newsavebox{\@brx}
\newcommand{\llangle}[1][]{\savebox{\@brx}{\(\m@th{#1\langle}\)}%
	\mathopen{\copy\@brx\kern-0.5\wd\@brx\usebox{\@brx}}}
\newcommand{\rrangle}[1][]{\savebox{\@brx}{\(\m@th{#1\rangle}\)}%
	\mathclose{\copy\@brx\kern-0.5\wd\@brx\usebox{\@brx}}}

\makeatletter
\def\maketitle{
	\@author@finish
	\title@column\titleblock@produce
	\suppressfloats[t]}
\makeatother

\newcommand{\w}{\bm{w}}

\newcommand{\cD}{\mathcal{D}}
\newcommand{\cL}{\mathcal{L}}

\newcommand{\cR}{\mathcal{R}}
\newcommand{\dd}{\mathrm{d}}
\newcommand{\sign}{\mathrm{sign}}
\newcommand{\Tr}{\mathrm{Tr}}

\DeclareMathOperator*{\argmax}{argmax}

\allowdisplaybreaks

\newcommand{\enr}[1]{{\color{red} #1}}

\makeatother

\begin{document}
	\title{A Flatness-Generalization Relation in the Teacher-Student Tree-Committee Machine}
	\date{\today}
	
	\author{Brandon Livio Annesi}
	\affiliation{Department of Quantitative Life Sciences, ICTP Trieste, Italy}
	
	\author{Davide Straziota}
	\thanks{This work was conducted prior to the author's employment at Amazon and is unrelated to his Amazon role.}
	\affiliation{Department of Computing Sciences, Bocconi University, 20136 Milano, Italy}

	\author{Enrico M. Malatesta}
	\affiliation{Department of Computing Sciences, Bocconi University, 20136 Milano, Italy}
	\affiliation{Bocconi Institute for Data Science and Analytics (BIDSA), Bocconi University, 20136 Milano, Italy}

	\begin{abstract}
	The flatness of the loss landscape at a minimizer is a widely used heuristic for reasoning about neural-network generalization, yet evidence for this relation is mostly empirical and controversial. We study this relation in a teacher-student tree committee machine, where both the ERM estimator and the Hessian spectrum are analytically tractable in the proportional high-dimensional limit. First, we use a zero-temperature Gibbs formulation to obtain predictions for the observables of the typical minimizers of the empirical loss. Secondly, we use Edwards-Jones formalism to derive the limiting Hessian resolvent around these typical minimizers. All predictions agree with finite-size gradient-descent simulations. Finally, we study three measures of flatness, namely the left and right edges and the spectral mean, and check if a decrease in generalization error as the dataset size is increased corresponds to an increase in flatness. We find that the answer strongly depends on the learning task and on the ratio of the number of parameters to the number of data points. In regression, the spectral mean and right edge correlate with the generalization error, while the left edge does so only in the overparametrized regime. In classification this correlation reliably holds only in the highly overparametrized phase, while for underparametrized networks it can even reverse.
	\end{abstract}

	\maketitle

	\section{Introduction}
	
	Why do overparametrized neural networks generalize after fitting high-dimensional
	training data? Among the many theoretical approaches that theoreticians have proposed to answer this question, a classic one has been to study the geometry of the loss landscape, in particular relating the generalization capacity of a network to the flatness of the minimizer \citep{hochreiter1997flat, mackay1992practical,hinton1993keeping, hochreiter1994simplifying}. The simple and heuristic argument is that if the difference between the population and empirical loss is not too big, then a function that minimizes the empirical loss and is also flat should also perform well on the population. This argument has been supported by many empirical works. For example \citep{keskar} show that increasing the batch size of SGD leads to both worse generalization and sharper minima. Furthermore, the notion of flatness has been used to propose new optimization algorithms that favor flat solutions, which typically perform better in terms of generalization \citep{foret2021sam, entropysgd}.
	
	Although this flatness-generalization relation is deeply rooted in machine learning lore, many subsequent works have found settings where this relation is broken \citep{zhang2021flatness,shoham2025flatness, andriushchenko2023modern}. For example \citep{dinh2017sharp} show that fully connected networks with ReLU activations can be reparametrized in a way that keeps the output invariant but varies many Hessian-based measures of flatness, thus potentially breaking this relation for any such measure. In an attempt to restore the relation, many works have come up with alternative measures of flatness that are invariant to this rescaling \citep{tsuzuku2020normalized,shoham2025flatness, petzka2020relative, pittorino2022deep, kwon2021asam}. Another line of research has compared the flatness of all minimizers of a loss, showing in some cases that minimizers with bad generalization properties can never be the flattest, even under reparametrization \citep{vardhan2026flatness, ding2024flat}. 
	
	A widely used tool for quantifying local flatness is the Hessian of the empirical loss. Its largest eigenvalue measures the curvature along the sharpest direction and arises from a quadratic approximation to worst-case neighborhood-based sharpness, whereas its trace measures the average curvature over parameter-space directions. Early empirical investigations found that the Hessian spectra of trained neural networks typically consist of a highly degenerate bulk concentrated near zero together with a small number of isolated eigenvalues \citep{sagun2017empirical, baldassi2020shaping}. Subsequent work developed scalable methods for estimating the entire spectrum and studied its evolution during training, revealing that the gradient can become strongly aligned with the eigenspaces associated with the largest eigenvalues \citep{ghorbani2019hessian}. Other studies connected these spectral structures to batch size and robustness \citep{yao2018hessian}, and showed that outliers and secondary bulks can encode class and cross-class structure in the data \citep{papyan2020traces, sabanayagam2023unveiling}. A unified phenomenological model proposed by \citep{fort2019emergent} connected several of these observations: the emergence of a small number of directions with large positive curvature, the alignment of the gradient with the corresponding eigenspace, and the non-monotonic evolution of the largest Hessian eigenvalue during training. \\
	
	On the theoretical side, random-matrix methods have been used to characterize the Hessian spectrum of simple neural network models \citep{pennington2017geometry,liao2021hessian, franz2015universal, asgari2025localminima, montanari2026topological}. In the case of networks with quadratic activations, related analyses have shown that spectral  transitions of the Hessian can mark changes in the stability or informativeness of stationary points \citep{mannelli2020complex,bonnaire2024timedependent,annesi2025bbp}. However, in none of these works has the Hessian been studied from a flatness-generalization perspective.
	
	The advantage of these models is that they offer a controlled setting in which macroscopic observables can be characterized analytically. For example, in teacher-student settings, data is labeled by a teacher network and used to train a student, so that generalization is determined by a small number of teacher--student and student--student overlaps. Statistical-mechanics methods have long been used in this setting to derive learning curves, phase transitions, and asymptotically exact dynamical equations \citep{watkin1993statmech,engel2001statmech,biehl1995online,saad1995soft}. Among the architectures studied in this framework, committee machines are particularly useful because they combine analytical tractability with important features of more complex models, including non-convexity and hidden-unit specialization. Both classical and modern works have used the replica method to uncover interesting learning phenomena such as SAT/UNSAT transitions \citep{barkai1992broken, barkai1990statistical, engel1992storage, annesi2025exact, Zavatone2021,baldassi2019properties}, specialization \citep{schwarze1992statistical,biehl1998phase,ahr1999statistical,barbier2025generalization}, and statistical-to-computational gaps \citep{aubin2018committee}. 
	
	\paragraph{Contributions.}
	In this work, we investigate the flatness–generalization relation in teacher–student tree committee machines in the proportional high-dimensional limit for two different learning tasks, regression and classification. We first use the replica method to characterize typical minimizers of the loss and derive analytical predictions for the order parameters that describe them. Depending on the learning task, the dataset size and dimensionality, and the regularization strength, we find that the solution is either replica symmetric (RS) or exhibits full replica symmetry breaking (FRSB). Intuitively, this means that in the first case the set of minimizers is connected, while in the latter, the low-loss configurations are organized into a continuous hierarchy of clusters characterized by different mutual overlaps, as in Parisi’s solution of mean-field spin glasses \citep{mezard1987spin}. Although we are able to solve the FRSB equations for a set of parameters of the model, we then approximate it with a 1RSB solution, which is numerically easier to handle, and which we confirm to be a good approximation of FRSB observables~\citep{annesi2025exact}.
	
	Secondly, we use the Edwards-Jones replica formalism \citep{edwards1976eigenvalue} to derive the spectrum of the Hessian matrix calculated on these minimizers. For square loss the resolvent is obtained from a scalar cubic equation whose coefficients are functions of the same order parameters that describe learning.
	From this, we calculate a set of scalar observables which are commonly used in the literature to quantify the flatness of a minimum, mainly the edges of the spectral distribution and its mean. Then, by varying dataset size, we check if a flatness-generalization correlation holds for some of these observables.\\
	We find that whether this relation holds or not is highly dependent on the observable, the learning task, and the parameters. For regression, we find that the relation holds both for the right edge (known in the literature as sharpness) and for the trace, while it holds for the left edge only in the overparametrized phase. For classification instead, we find that it holds for all three observables only in the overparametrized phase: in all other phases the relation is broken, and in some cases flatness can also anticorrelate with generalization.
	
	Finally, we validate all our theoretical predictions with an extensive set of gradient descent simulations. We find good agreement with the theory even in the replica symmetry broken regime, where the exact equilibrium sampling is known to be challenging for stable algorithms in related mean field spin glass models \cite{el2022sampling}.
	

	\section{The Model}
	\label{sec:model}
	
	\begin{figure}[t]
		\centering
		\includegraphics[width=0.7\textwidth]{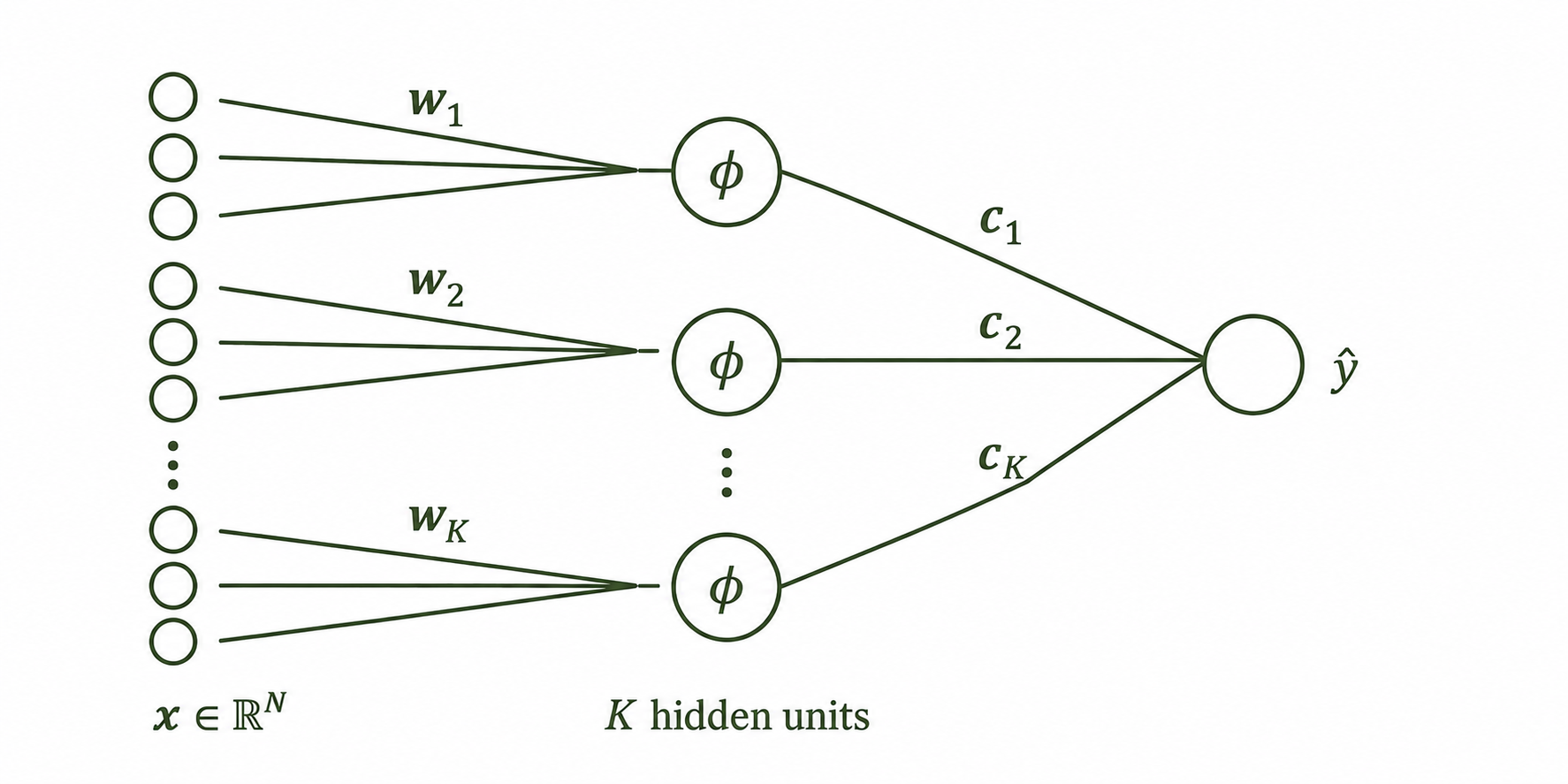}
		\caption{Schematic representation of a tree committee machine.}
		\label{fig:committee}
	\end{figure}

	We consider a teacher--student learning problem where both the student and teacher are tree-committee machines. This is a two-layer neural network where the $K$ hidden units have non-overlapping receptive fields, and the only trainable parameters are the first layer weights $\{\boldsymbol{w}_l\}_{l=1}^K$, with $\boldsymbol{w}_l\in\mathbb{R}^{N/K}$ (a schematic view of this architecture is shown in Figure \ref{fig:committee}). The training dataset consists of \(P=\alpha N\) input--label pairs $\mathcal{D} = \{(\boldsymbol{x}^\mu,y^\mu)\}_{\mu=1}^P$, where we take the inputs \(x^\mu_{i}\sim\mathcal{N}(0,1)\). For any choice of first-layer weights \(\boldsymbol{w}\), we define the committee preactivation on pattern \(\mu\) as
	\begin{equation}
		z^\mu(\boldsymbol{w})
		\equiv
		\frac{1}{\sqrt{K}}\sum_{l=1}^K c_l\,
		\varphi\!\left(
		\sqrt{\frac{K}{N}}
		\sum_{i=1}^{N/K} w_{li}\,x^\mu_{li}
		\right),
		\label{eq:committee_preactivation}
	\end{equation}
	where we have used the indexing convention \(x_{li}^\mu = x^\mu_{i+(l-1)N/K}\).
	The unknown teacher first-layer weights are drawn from a Gaussian distribution $w_{li}^\star \sim \mathcal{N}(0,1)$, the known second layer weights are chosen such that $\sum_{l=1}^K c _l = 0$ and $\sum_{l=1}^K c_l^2 =K$, and the output function depends on the task: for regression it is simply $f_\star(z)=z$, while for binary classification it is $f_\star(z)=\sign(z)$. The labels are generated by the teacher network, while the student has the same architecture and has to learn the first layer weights:
	\begin{equation}
		y^\mu=f_\star\!\left(z^\mu(\boldsymbol{w}^\star)\right),
		\qquad
		\hat y^\mu(\boldsymbol{w})=z^\mu(\boldsymbol{w}).
		\label{eq:teacher_student_outputs}
	\end{equation}
	Note how the student shares the same instance of the second layer weights as the teacher.

	The student learns the weights $\boldsymbol{w}$ by minimizing the \(\ell_2\)-regularized empirical loss
	\begin{equation}
		\cL(\w)
		=
		\frac{1}{\alpha}\sum_{\mu=1}^{P}
		\ell\!\left(y^\mu,\hat y^\mu(\w)\right)
		+
		\frac{\lambda}{2}\|\w\|^2 .
		\label{eq:loss_main}
	\end{equation}
	
	The factor \(1/\alpha\) ensures that the data-fitting and regularization terms are both of order \(N\) as \(\alpha\) is varied. We define the set of regularized empirical risk minimizers (ERM) by
	\begin{equation}
		\widehat{\w}_{\mathrm{ERM}}
		\in
		\operatorname*{arg\,min}_{\w}\cL(\w).
		\label{eq:erm_estimator}
	\end{equation}
	To study its properties analytically, we introduce the Gibbs measure
	\begin{equation}
		p_\beta(\w\mid\cD)
		=
		\frac{1}{Z_\beta}
		\exp[-\beta\cL(\w)],
		\qquad
		Z_\beta
		=
		\int d\w\,\exp[-\beta\cL(\w)].
		\label{eq:gibbs_main}
	\end{equation}
	In the $\beta\to\infty$ limit, this measure concentrates on the set of ERM estimators. 
	
	The Gibbs formulation of the ERM~\eqref{eq:gibbs_main} can be used also to access its generalization ability. For a fresh Gaussian input \(\boldsymbol{x}_\star\), let \(y_\star\) be the teacher label and \(\hat y_\star(\boldsymbol{w})\) the student's preactivation. In regression we use the test mean-squared error
	\begin{equation}
		\epsilon_g^{\rm regr}
		=
		\mathbb{E}_{\boldsymbol{x}_\star}
		\mathbb{E}_{\mathcal{D}}
		\frac{1}{2}
		\left\langle
		\left(y_\star-\hat y_\star(\boldsymbol{w})\right)^2
		\right\rangle_{\boldsymbol{w}\mid\mathcal{D}}\,.
		\label{eq:generalization_regression_def}
	\end{equation}
	Here \(\langle \cdot\rangle_{\boldsymbol{w}\mid\mathcal{D}}\) denotes the average over the Gibbs measure~\eqref{eq:gibbs_main}. 
	In classification we use the probability that teacher and student disagree
	\begin{equation}
		\epsilon_g^{\rm class}
		=
		\mathbb{E}_{\boldsymbol{x}_\star}
		\mathbb{E}_{\mathcal{D}}
		\left\langle
		\Theta\!\left(-y_\star\hat y_\star(\boldsymbol{w})\right)
		\right\rangle_{\boldsymbol{w}\mid\mathcal{D}}.
		\label{eq:generalization_classification_def}
	\end{equation}
	
	The next section describes how the set of ERM estimators can be characterized by a finite set of macroscopic order parameters.
	\section{The ERM Estimator}
	\label{sec:states}
	
	\subsection{Replica description and generalization error}
	\label{subsec:replica_states}
	
	We characterize typical ERM estimators in the proportional limit \(N,P\to\infty\) with \(\alpha=P/N\) fixed. The hidden-unit number is kept fixed while this thermodynamic limit is taken, and only afterwards we send \(K\to\infty\). In this regime the quantity $\log Z_\beta / N$ concentrates, so its typical (or most probable) value is given by the quenched average, which is called the free entropy density 
	\begin{equation}
		\phi_\beta
		=
		\lim_{N\to\infty}
		\frac{1}{N}\,
		\mathbb{E}_{\mathcal{D}} \log Z_\beta,
		\label{eq:free_entropy_main}
	\end{equation}
	Here the notation $\mathbb{E}_{\mathcal{D}}$ denotes the average over the training set that depends on the particular realization of the inputs $\{\boldsymbol{x}^\mu\}_{\mu = 1}^{\alpha N}$ and the teacher weights $\boldsymbol{w}^\star$. 
	
	\begin{figure}[t]
		\centering
		\includegraphics[width=\textwidth]{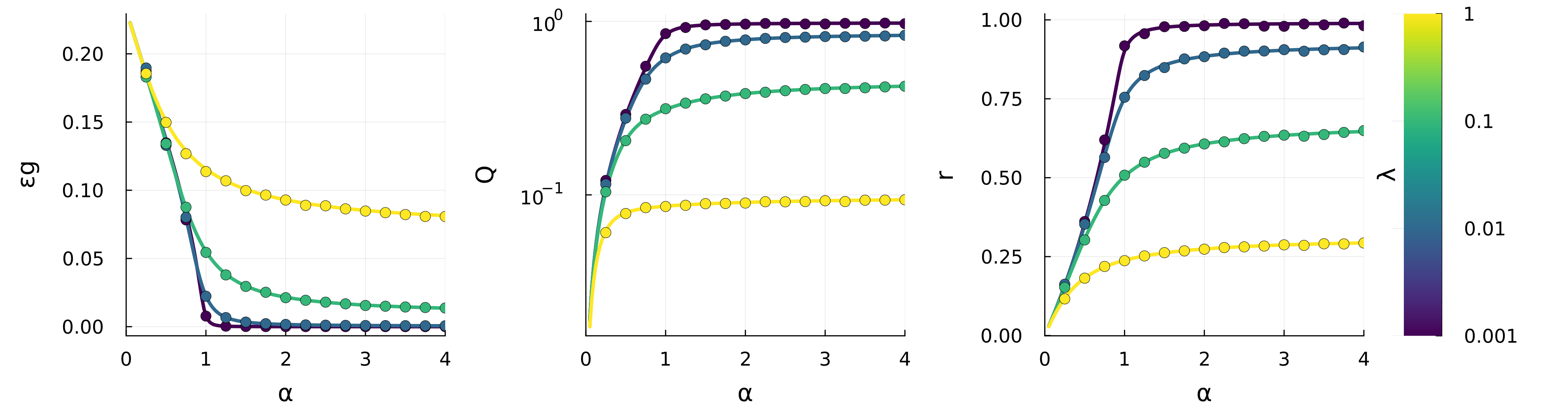}
		\caption{Generalization error and order parameters for regression. The RS theory predicts the test error
			$\epsilon_g$, squared norm $Q$, and teacher--student overlap $r$ as functions
			of sample complexity $\alpha$ and regularization strength $\lambda$. Curves
			are theoretical predictions; markers report finite-size ($N=2000,K=20$) GD simulations, averaged over 20 seeds. Error bars are not shown because they are very small.}
		\label{fig::equilibrium_regression}
	\end{figure}
	
	We evaluate the free entropy using the replica method~\citep{mezard1987spin}, which expresses it in terms of a finite set of order parameters.
	These order parameters have a direct physical interpretation: they describe, after averaging over the dataset $\mathcal{D}$, the typical mutual overlap between independently sampled weights from $p_\beta(\cdot | \mathcal{D})$ and their typical alignment with the teacher. For hidden unit \(l\), let us define:
	\begin{equation}
		Q_l^{a}
		=
		\mathbb{E}_{\mathcal{D}}
		\frac{K}{N}
		\sum_{i=1}^{N/K}(w^a_{li})^2, \qquad
		q_l^{ab}
		=
		\mathbb{E}_{\mathcal{D}}
		\frac{K}{N}
		\sum_{i=1}^{N/K}w^a_{li}w^b_{li}
		\,, \qquad
		r^a_l
		=
		\mathbb{E}_{\mathcal{D}}
		\frac{K}{N}
		\sum_{i=1}^{N/K}w^a_{li}w^\star_{li}
		.
		\label{eq:physical_order_parameters}
	\end{equation}
	where $\boldsymbol{w}^a,\boldsymbol{w}^b\sim p_\beta(\cdot|\mathcal{D})$ are iid samples of the Gibbs distribution~\eqref{eq:gibbs_main}. Thus \(q_l^{ab}\) is the overlap between two Gibbs samples, \(Q^a_l\) is the squared norm of one sampled weight, and $r_l$ is the teacher-student alignment. Since the hidden units are statistically equivalent and their receptive fields involve disjoint, statistically independent subsets of the inputs, we can drop the hidden-unit index $l$: $q_l^{ab} = q^{ab}$ and $r_l=r$.
	
	Although these quantities are averaged over $\mathcal{D}$, they are still random variables that depend on the draws from the Gibbs distribution, and could therefore have a non-trivial distribution. While it can be shown that both the norm and the teacher-student overlap have a trivial delta function distribution, this does not necessarily happen for the overlap $q^{ab}$. The remaining choice is how to approximate the distribution $P(q)$ of this quantity. 
	
	Under the replica symmetric (RS) ansatz this distribution is a single delta peak centered around a quantity $q_0$: $P(q) = \delta(q - q_0)$. In other words, all pairs of independently sampled students have the same typical overlap $q^{ab}=q_0$. The overlaps $q^{ab}$ thus behave similarly to the norm \(Q\) and the teacher-student overlap \(r\).
	
	Replica symmetry breaking (RSB) ansatzes allow the same distribution to have more complicated shapes. The simplest one corresponds to the one-step replica symmetry breaking (1RSB) ansatz, which parametrizes the overlap distribution by two values $q_1>q_0$:	$P(q) = m\,\delta(q-q_0) + (1-m)\,\delta(q-q_1)$. 
	This can be interpreted by imagining samples that are organized into clusters: two samples in the same cluster have overlap \(q_1\), while samples in different clusters have overlap \(q_0\). The Parisi parameter \(m\) encodes the probability of observing the inter-cluster overlap \(q_0\). 
	More elaborate RSB ansatzes introduce additional overlap scales, and in the full-RSB limit the discrete set of peaks becomes a continuous probability distribution over overlaps. We discuss in the next subsection which of these ansatzes is selected, depending on the task and the particular region of the phase diagram examined. The value of the order parameters that parametrize the probability distributions is then obtained by solving a set of saddle point equations; see discussion in appendix~\ref{sec::Equilibrium}.
	
	\begin{figure}[t]
		\centering
		\includegraphics[width=\textwidth]{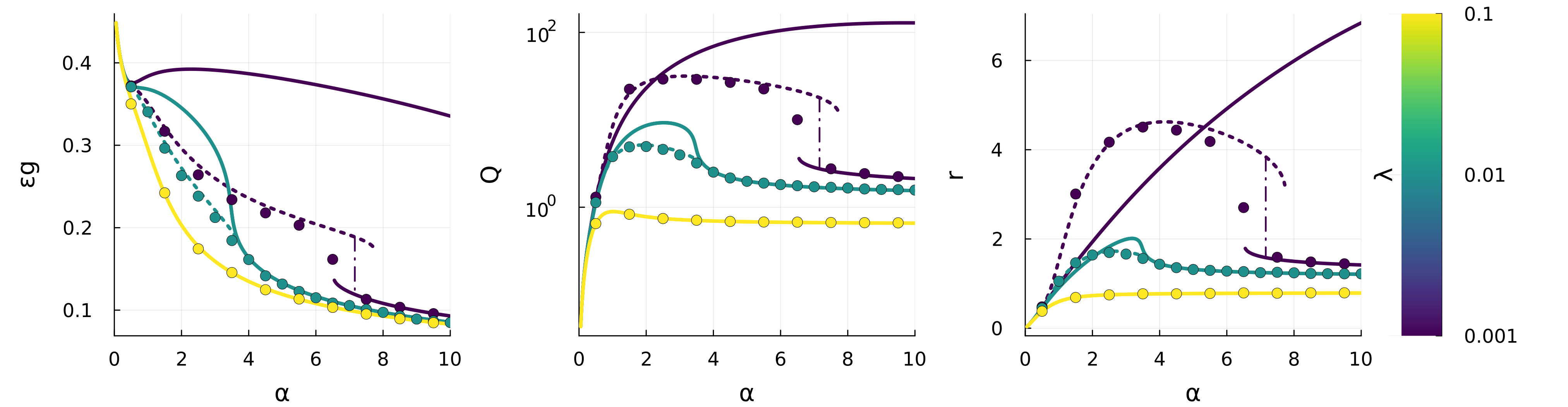}
		\caption{Classification order parameters. Solid curves denote RS solutions, while dashed curves denote 1RSB solutions. Markers report finite-size GD
			simulations for $N=2000,K=20$, averaged over 20 seeds. Error bars are not shown because they are very small.}
		\label{fig:classification_phase}
	\end{figure} 
	
	From these order parameters we can also access the generalization performance of the network. In particular, the definitions in Eqs.~\eqref{eq:generalization_regression_def} and \eqref{eq:generalization_classification_def} can be evaluated from the joint Gaussian law of the teacher and student outputs on a fresh test input. In the \(K\to\infty\) limit these errors become explicit functions of the order parameters $Q$ and $r$:
	\begin{subequations}
		\label{eq:generalization_errors_main}
		\begin{align}
			\epsilon_g^{\rm class} &=
			\frac{1}{\pi}
			\arccos\left(
			\frac{D_0}{\sqrt{\Phi_\star\left(\Phi(Q)-\Phi(0)\right)}}
			\right), \\
			\epsilon_g^{\rm regr} &=
			\frac{\Phi(Q)-\Phi(0)+\Phi_\star-2D_0}{2}.
		\end{align}
	\end{subequations}
	The quantities $\Phi_\star$, $D_0$ and the function $\Phi(\bullet)$ appearing in \eqref{eq:generalization_errors_main} are written in terms of the infinite-width NNGP kernel
	\begin{equation}
		\mathcal{K}(Q_\star,Q,q)
		\equiv
		\int Dx\,Dy\,
		\varphi(\sqrt{Q_\star}x)\,
		\varphi\left(
		\frac{q}{\sqrt{Q_\star}}x
		+
		\sqrt{Q-\frac{q^2}{Q_\star}}y
		\right),
		\label{eq:nngp_kernel_main}
	\end{equation}
	where \(Dx\) and \(Dy\) denote standard Gaussian measures. We use the shorthand \(\Phi_\star=\mathcal{K}(1,1,1)-\mathcal{K}(1,1,0)\), \(D_0=\mathcal{K}(1,Q,r)-\mathcal{K}(1,Q,0)\), and \(\Phi(q)=\mathcal{K}(Q,Q,q)\). 
	
	In the rest of the paper, we will focus on the square loss function
	\begin{equation}
		\label{eq::square_loss}
		\ell(y,\hat y)=\frac{1}{2}(y-\hat y)^2.
	\end{equation}
	Although for classification tasks other losses may achieve better performance, the square loss allows a Gaussian integral entering the replica calculation to be evaluated analytically (see appendix~\ref{sec::Equilibrium} for additional detail). It therefore gives a tractable high-dimensional landscape in which to study the properties of the ERM estimator. For a generic loss, the same integral must instead be evaluated numerically. We will also choose $\varphi(x)=\text{erf}(x)$ since the NNGP kernel in~\eqref{eq:nngp_kernel_main} can be evaluated analytically.

	\subsection{Regression: a replica symmetric phase diagram}
	
	First of all, we analyze the properties of the ERM estimator in the regression task, where the teacher output function is the identity. In this case, teacher and student architectures are fully matched, and we find that the full phase diagram is captured by a Replica Symmetric ansatz.  Let us stress that a priori there is no theoretical guarantee that this should be the case: Replica Symmetry is only guaranteed for convex losses or in Bayesian-optimal settings, neither of which applies here.
	
	The test error, norm and overlap with the teacher are shown in Figure \ref{fig::equilibrium_regression} as a function of $\alpha$ for various values of $\lambda$. The behavior of the estimator is very similar to the Ridge Regression estimator: as long as the regularization is low enough, for $\alpha>1$ the test error drops close to zero, and the teacher is almost fully recovered. For $\alpha <1$ we find that the test error is non-monotonic in $\lambda$ (not shown in the plot), while for $\alpha>1$ it is monotonically decreasing with decreasing $\lambda$. \\
	We then validate this RS ansatz by performing Gradient Descent experiments, and obtain an excellent match with our theory. More information on the numerics can be found in the appendix \ref{app:numerics}.

	\subsection{Classification: Replica Symmetry Breaking}
	
	In the classification task, the ERM estimator displays a much richer phase diagram than its regression counterpart. The test error and the associated order parameters are shown in Fig.~\ref{fig:classification_phase}.
	At sufficiently low sample complexity, the system is well described by a replica symmetric (RS) saddle point for any value of the regularization strength $\lambda$.
	
	As $\alpha$ increases, however, the shape of the landscape changes. For sufficiently small $\lambda$, an intermediate window of $\alpha$ opens in which the correct equilibrium description requires breaking the replica symmetry. Although the exact description of this phase involves full replica symmetry breaking (see Appendix~\ref{sec::kRSB} for detail), we use a 1RSB approximation, which captures the relevant observables with only small corrections from higher RSB levels. 
	
	Finally, once $\alpha$ exceeds a threshold that depends explicitly on $\lambda$, the system exits the RSB phase and returns to a replica symmetric description, but of a qualitatively different character than the one found at small $\alpha$. We refer to this final regime as the \emph{recovery phase}. Here the test error drops sharply and the teacher-student cosine similarity $r/\sqrt{Q}$ increases markedly, indicating that the Gibbs measure has become dominated by a single, well-recovered minimum with substantially better generalization properties than the states populating the RSB region. The character of the transition connecting the RSB phase to the recovery phase itself depends sensitively on $\lambda$: for sufficiently small regularization (for instance $\lambda = 0.001$ in Fig.~\ref{fig:classification_phase}) the transition is discontinuous, with the order parameters jumping abruptly between the RSB branch and the recovery-phase branch, consistent with a first-order transition and an associated coexistence region. For larger values of $\lambda$ the transition instead becomes continuous, with the RSB branch merging smoothly into the recovery phase as $\alpha$ is increased.

	\section{Hessian Spectrum in Minimizers}
	\label{sec:hessian}
	
	\begin{figure}[t]
		\centering
		\includegraphics[width=\textwidth]{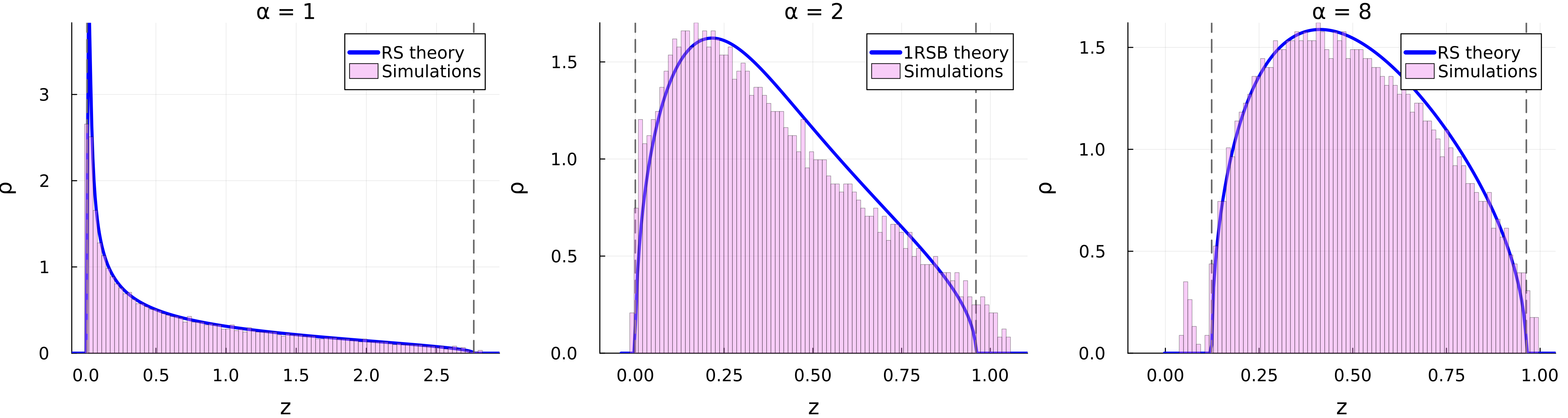}
		\caption{Empirical spectrum vs theoretical spectrum for $N=2000, K=20$ and $\lambda = 0.01$. The left panel shows regression; the middle and right panels show classification.}
		\label{fig:Hessian_spectrum}
	\end{figure}
	
	Having described the macroscopic properties of typical minimizers, we turn to computing their local curvature. Let $\mathcal{H}$ denote the Hessian of the data term in \eqref{eq:loss_main} evaluated at a typical trained configuration. 
	\begin{equation}
		\mathcal{H} 
		=
		\frac{1}{\alpha}\sum_{\mu=1}^{P}  \frac{\partial^2 \ell\!\left(y^\mu,\hat y^\mu(\w)\right)}{\partial \boldsymbol{w} \partial \boldsymbol{w}^T}
	\end{equation}
	We are interested in computing the limiting spectral distribution of this matrix, $\rho_\mathcal{H}(z) = \lim_{N\to\infty} \frac{1}{N}\sum_{i=1}^N \delta(z-\lambda_i)$ where $\{\lambda_i\}_{i=1}^N$ are its eigenvalues. To this end we compute the Stieltjes transform $\cR_\mathcal{H}(z)= \lim_{N\to\infty}\frac{1}{N}\Tr\,(zI-\mathcal{H})^{-1}$ from which the limiting spectral distribution can be obtained via the Stieltjes inversion formula
	\begin{equation}
		\rho_\mathcal{H}(z)
		=
		\lim_{\eta\to 0}\frac{1}{\pi}
		\mathrm{Im}\,\cR_\mathcal{H}(z-i\eta).
		\label{eq:stieltjes_main}
	\end{equation}
	The Stieltjes transform can in turn be written as
	\begin{equation}
		\mathcal{R}_\mathcal{H}(z) = -\lim_{N\to\infty}\frac{2}{N} \frac{\partial}{\partial z} \ln \det(\mathcal{H} - zI)^{-1/2}
	\end{equation}
	which, using a Gaussian integral representation of the determinant, allows us to use the replica method \citep{edwards1976eigenvalue}, see Appendix~\ref{sec::Hessian_computation} for additional detail.

	In the case of the square loss function~\eqref{eq::square_loss}, the Stieltjes transform satisfies a cubic equation
	\begin{equation}
		1+a_1(z)\cR_{\mathcal{H}}+a_2(z)\cR_{\mathcal{H}}^2+a_3\cR_{\mathcal{H}}^3=0,
		\label{eq:cubic_main}
	\end{equation}
	where the coefficients $a_1(z),a_2(z),a_3$ are explicit functions of $\alpha$, the activation kernel and its derivatives, and the order parameters and can be found in Appendix~\ref{sec::Hessian_square_loss}. Consequently, the Hessian spectrum is determined by the same macroscopic state variables that were calculated above.

	In Figure \ref{fig:Hessian_spectrum} we compare the theoretical spectrum with finite-$N$ Gradient Descent simulations. The agreement is excellent in the RS phases, while a small discrepancy remains in the detailed spectral shape in the RSB phase. Since higher-order RSB corrections to the equilibrium order parameters are small (see Appendix~\ref{sec::kRSB}), we expect this mismatch to be more likely related to the extremely slow equilibration times characteristic of full-RSB landscapes~\citep{sompolinsky1981dynamic,sompolinsky1982relaxational,cugliandolo1994out}. 

	
	
	Finally, note that in the high-$\alpha$ recovery phase, finite-$N$ simulations show a set of $K$ eigenvalues detaching from the left side of the continuous bulk (right panel of Fig. \ref{fig:Hessian_spectrum}), with corresponding eigenvectors highly aligned to the teacher network. These modes are not captured by our limiting spectral density, since the large $N$ limit is taken with fixed $K$ and hence their spectral weight is $K/N\to0$. We therefore exclude them when comparing the theoretical bulk edge with simulations, and report the edge of the continuous part of the spectrum. 
	We also note that those isolated Hessian eigenvalues can have important dynamical consequences in other models. In phase retrieval, for example, an outlier eigenvalue can be associated with an eigenvector aligned with the planted signal, and its appearance or disappearance through a BBP transition is closely tied to whether gradient descent can escape marginal minima and recover the signal~\citep{bonnaire2024timedependent}.

	\begin{figure}[t]
		\centering
		\includegraphics[width=\textwidth]{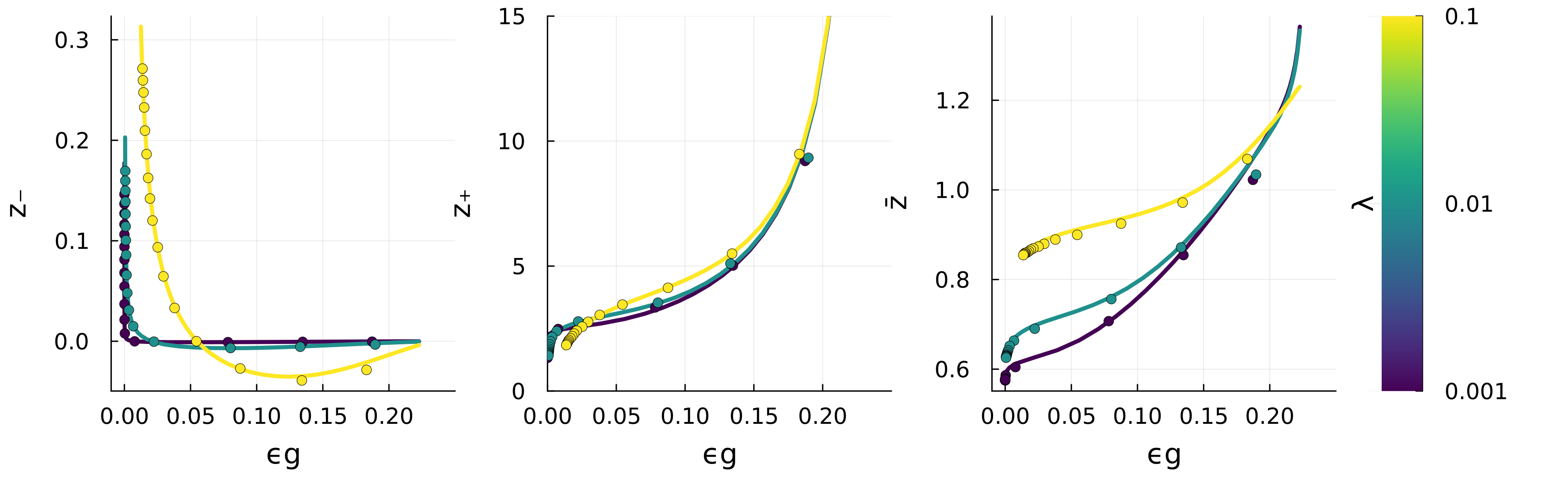}
		\caption{Flatness observables versus generalization error in regression. Markers are given by GD simulations with $N=2000,K=20$. In the phases where the empirical spectrum develops outliers, we show the smallest eigenvalue of the continuous bulk.}
		\label{fig:flatness_regression}
	\end{figure}
	
	\section{Flatness is phase-dependent}
	\label{sec:flatness}
	
	We now ask which Hessian observables track generalization. We focus on three basic spectral summaries: the mean curvature
	\begin{equation}
		\bar z = \int x\,\rho_{\mathcal{H}}(x)\,\dd x = \frac1N\Tr \, \mathcal{H},
	\end{equation}
	and the left and right spectral edges $z_-$ and $z_+$. The right edge $z_+$ is a sharpness proxy, the trace measures the average curvature scale, and the left edge probes soft or marginal directions. Let us point out that our model possesses neither reparametrisation nor permutation symmetries, and therefore these flatness measures are not subject to the inconsistencies pointed out by \cite{dinh2017sharp}. As we vary $\alpha$, the generalization properties of the minimizers change. To check if a flatness-generalization relation holds in this setting, we plot these spectral observables as a function of the generalization error as $\alpha$ is varied, and visually inspect if a positive correlation holds. 
	
	\paragraph{Regression.}
	We first address this question in the regression task. Figure~\ref{fig:flatness_regression} shows the relation between test error and the three flatness observables. Across the range of regularization strengths that we explored, the spectral mean and the right edge are positively correlated with the test loss. In other words, both the average curvature and the largest curvature provide a geometric quantity that correlates with generalization. \\
	The left edge behaves differently. It correlates with the test error only in the overparametrized regime (where $\alpha < 1)$ and with a degree that depends on the regularization $\lambda$, higher $\lambda$ giving a higher correlation. In the underparametrized phase, instead, this relation breaks down, the left edge becoming anticorrelated with the test error.

\begin{figure}[t]
		\centering
		\includegraphics[width=\textwidth]{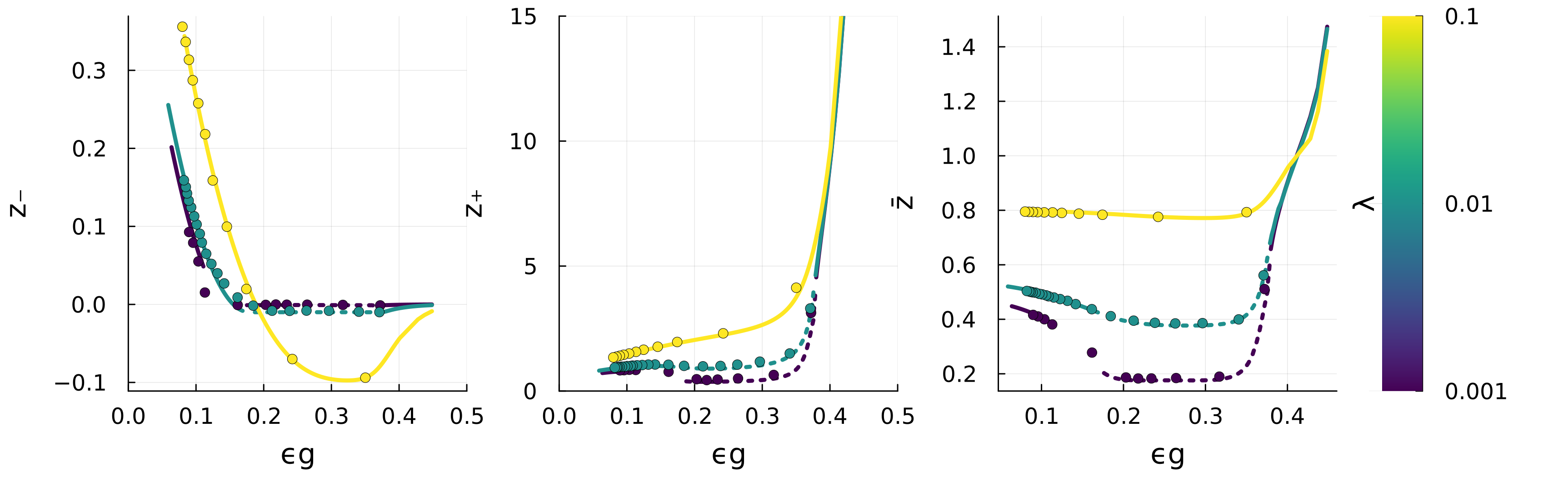}
		\caption{Flatness observables versus generalization error in classification. Solid curves are RS branches and dashed ones are 1RSB branches. Markers are given by GD simulations with $N=2000,K=20$. In the phases where the empirical spectrum develops outliers, we show the smallest eigenvalue of the continuous bulk.}
		\label{fig:flatness_classification}
	\end{figure}

	\paragraph{Classification.}
	A similar analysis can be performed for classification, using the 1RSB solution as the best available description of the relevant minimizers. In Fig.~\ref{fig:flatness_classification} we show the relation between spectral observables and test error for three values of the regularization $\lambda$, one for which the transition from the RSB phase to the recovery RS phase is continuous, and one for which it is discontinuous, and one for which the full phase diagram is RS. The resulting picture is qualitatively consistent with regression only within the first overparametrized RS phase, where both the spectral mean and the left and right edges remain reliably correlated with generalization. The correspondence breaks down in the other two phases: in the RSB regime, spectral observables are essentially uncorrelated with the generalization error, while in the recovery phase the correlation can in some cases reverse sign, with flatter minima associated with worse generalization. Hessian-based flatness measures are therefore phase-dependent diagnostics in classification, rather than universal proxies for predictive performance.
	
	An interesting remark is that the left edge in the RSB phase is constant and equal to $-\lambda$. Indeed, the left edge cannot go below this value, since we are studying the Hessian of minimizers, and therefore the eigenvalues of the Hessian of the full loss (which contains a $+\lambda$ shift compared to the Hessian we are studying) cannot be negative. Our observation is thus that the RSB states predicted by our theory are marginal, a prediction which is matched by our simulations.
	
	\section{Discussion and limitations}
	\label{sec:discussion}
	
	We have studied Hessian-based flatness measures of empirical-risk minimizers in a teacher-student tree committee machine model. 
	The ERM estimators are characterized by a finite set of order parameters, which determine both their learning performance and the limiting Hessian spectrum around the learned solution. We focused on three standard flatness measures: the spectral mean, which captures the average curvature, and the left and right spectral edges, which probe the softest and sharpest directions. We validated the resulting predictions against finite-size gradient-descent simulations, finding quantitative agreement for all observables considered. %
	
	The central conclusion is that the flatness–generalization relation is phase-dependent rather than universal. In regression, where the phase diagram is replica symmetric, the spectral mean and right edge correlate with the test error across the regularizations we consider, while the left edge tracks it only for $\alpha<1$. In classification the same observables predict generalization only in the overparametrized phase; in the intermediate RSB phase they are essentially uncorrelated with it, and in the high $\alpha$ replica symmetric recovery phase the correlation can reverse. 
	
	Two features of the RSB phase are worth isolating. First, the left edge is pinned at $-\lambda$ throughout this phase. Since the full regularized loss is minimized, its Hessian is positive semidefinite, so the data-term Hessian cannot have eigenvalues below $-\lambda$; our solution saturates this bound, indicating that these minimizers are marginal. This is reproduced by the simulations. Second, at large $\alpha$ the empirical Hessian develops a finite number of outlier eigenvalues to the left of the bulk. These are invisible to the limiting spectral density, which sees only $O(N)$ eigenvalues
	; their origin, and their relation to the class-structured Hessian outliers reported in deep classifiers, lies outside the present framework.
	
	Finally, the context in which we analyze the flatness-generalization relation is slightly different from the one that is typically found in empirical works: there the relation is observed with a fixed dataset among different minimizers of the same loss landscape. In our setting, instead we modify the loss landscape by varying the dataset size to input dimension ratio. Strictly speaking, our results concern how curvature and generalization co-vary, and need not transfer directly to the fixed-data setting in which the relation is usually probed. Nevertheless, studying this relation as the landscape is varied provides a complementary perspective on the interplay between curvature and generalization. Furthermore, varying $\alpha$ has the advantage of sweeping across the RS, RSB, and recovery phases, which exposes the phase dependence we report. A similar analysis at fixed $\alpha$, tracking distinct minimizers of a single landscape, would be needed to establish whether the same phase-conditioned picture holds within a landscape as well as across a family of them; we leave this to future work.

	\section{AI Disclosure}
	
	In this work, we used generative AI tools for small steps in the replica computation, which were then independently verified. Additionally, we used generative AI tools for some of the plots, and for polishing the final version of this manuscript. We have reviewed all AI-assisted work. We take responsibility for the final content of this work, including text, claims or artifacts produced with the aid of generative AI.
	
	\bibliography{references}

	\clearpage

	\title{A Flatness-Generalization Relation in the Teacher-Student Tree-Committee Machine \\ SUPPLEMENTAL INFORMATION}
	\maketitle
	
	\onecolumngrid 
	
	\makeatletter
	\patchcmd{\tableofcontents}{\@starttoc{toc}}{\thispagestyle{empty}\pagestyle{empty}\@starttoc{toc}}{}{}
	\makeatother
	
	\tableofcontents
	
	\appendix
	
	\section{Setting and objective} \label{sec::setting&objective}
	
	We will consider two identical architectures given by a one-hidden layer architecture with non-overlapping first layer weights (tree committee machines)
	\begin{subequations}
		\begin{align}
			y^\mu &= f_\star \left[ \frac{1}{\sqrt{K}} \sum_{l=1}^K c_l \, \varphi \left( \sqrt{\frac{K}{N}} \sum_{i=1}^{N/K} w_{li}^\star x_{li}^\mu \right) \right] \\
			\hat y^\mu(\boldsymbol{w}) &= \frac{1}{\sqrt{K}} \sum_{l=1}^K c_l \, \varphi \left( \sqrt{\frac{K}{N}} \sum_{i=1}^{N/K} w_{li} x_{li}^\mu \right) \equiv \frac{1}{\sqrt{K}} \sum_{l=1}^K c_l \, \varphi_l
		\end{align}
	\end{subequations}
	where $f_\star(x) = x$ for regression and $f_\star(x) = \mathrm{sign}(x)$ for classification. Moreover the $P=\alpha N$ input data $\boldsymbol{x}^\mu$ will be standard normal i.i.d. random variables as well as the weights of the teacher $w_{li}^\star$. The second layer weights $c_l$ can be considered either to be fixed or sampled from a distribution. In the following we will assume that they satisfy the following conditions
	\begin{subequations}
		\label{eq::conditions_c}
		\begin{align}
			\sum_l c_l &= 0 \\
			\sum_l c_l^2 &= K
		\end{align}
	\end{subequations}
	i.e. to have mean $0$ and variance $1$.
	We consider the empirical risk minimization
	\begin{equation}
		\hat{\boldsymbol{w}} = \mathrm{argmin}_{\boldsymbol{w}}\left[ \mathcal{L}(\boldsymbol{w}) \right] 
	\end{equation}
	of a loss function $\mathcal{L}$ given by
	\begin{equation}
		\mathcal{L}(\boldsymbol{w}) = \sum_\mu \ell(y^\mu, \hat y^\mu(\boldsymbol{w}))  + \frac{\lambda \alpha}{2} \lVert \boldsymbol{w} \rVert^2 \,.
		\label{eq::loss_appendix}
	\end{equation}
	$y^\mu$ and $\hat y^\mu$ are the output and the preactivation of the output of the teacher and student networks and $\lambda$ represents the L2 regularization of the weights of the network. The ERM estimator $\hat{\boldsymbol{w}}$ can be obtained by studying the large $\beta$ limit of the Gibbs measure (or posterior distribution)
	\begin{equation}
		\label{eq::Gibbs_measure}
		p(\boldsymbol{w} | \mathcal{D}) = \frac{e^{- \beta \mathcal{L}(\boldsymbol{w})}}{Z_\beta} 
	\end{equation}
	where we have denoted by $\mathcal{D} = \left\{ \boldsymbol{x}^\mu, y^\mu \right\}_{\mu=1}^{\alpha N}$ the dataset and by $Z_\beta$ the partition function 
	\begin{equation}
		\label{eq::partition_function_equilibrium}
		Z_\beta = \int d \boldsymbol{w} \, e^{- \beta \mathcal{L}(\boldsymbol{w})}
	\end{equation}
	Note that the loss defined here~\eqref{eq::loss_appendix} differs from the one given in the main text by a factor $\alpha$. This factor can be absorbed in the definition of the inverse temperature and does not give any contribution in the large $\beta$ limit on which we focus.

	\section{Equilibrium measure} \label{sec::Equilibrium}
	
	So far we did not specify how to compute the value of $q$, $r$ and $Q$. This can be done by studying the equilibrium measure induced by the partition function defined in~\eqref{eq::partition_function_equilibrium}, in particular via the corresponding equilibrium free entropy
	\begin{equation}
		\phi^{\mathrm{eq}} = \lim\limits_{N\to \infty}\frac{1}{N} \mathbb{E}_{\mathcal{D}} \ln Z_\beta \,.
	\end{equation}
	This quantity can be computed with standard methods, namely the replica method~\cite{mezard1987spin} which is based on the following identity
	\begin{equation}
		\ln Z_{\beta} = \lim\limits_{n\to 0} \partial_n Z^n_{\beta}
	\end{equation}
	We report here the outcome of the computation, which is reported elsewhere~\cite{baldassi2019properties}. In the following we will denote by $l$ the index that runs over the hidden units $l=1, \dots, K$, and by $a, b$ the indices running over the number of replicas $a, b = 1, \dots, n$. 
	One finds that the equilibrium free entropy is obtained by a saddle point computation over a set of \emph{order parameters} that can be conveniently condensed in $K$ symmetric matrices of dimension $(n+1)\times(n+1)$ that we will call $Q_l^{\alpha \beta}$. The indices $\alpha$, $\beta$ will be indexed by $\alpha, \beta = 0, \dots, n$ for convenience. This matrix has a block structure of the type
	\begin{equation}
		\label{eq::Q_blocks}
		\boldsymbol{Q}_l = \begin{pmatrix}
			1 & r^{a}_l \\
			r_l^a & q_l^{ab}
		\end{pmatrix}
	\end{equation}
	where $q_l^{ab} \equiv \frac{K}{N} \sum_{i=1}^{N/ K} w_{li}^a w_{li}^b$ with $a<b$ represents the overlap between the $l$-th hidden unit of two students extracted from the Gibbs measure~\eqref{eq::Gibbs_measure}, $q_l^{aa} = \frac{K}{N} \sum_{i=1}^{N/ K} (w_{li}^a)^2$ is the typical squared norm of students weights and finally $r_l^{a} \equiv \frac{K}{N} \sum_{i=1}^{N/ K} w_{li}^a w_{li}^\star$ is the overlap between the teacher and a student. Moreover $Q_l^{00} = 1$, which corresponds to the norm of the teacher, which is 1 in the large $N$ limit by the central limit theorem. 
	
	One finds
	\begin{subequations}
		\label{eq::equilibrium_before_ansatz}
		\begin{align}
			\phi^{\mathrm{eq}} &= \lim\limits_{n\to 0} \partial_n \; \mathrm{extr}_{\boldsymbol{Q}} \left[ G_{SI}^{\mathrm{eq}}(\boldsymbol{Q}) + \alpha G_E^{\mathrm{eq}}(\boldsymbol{Q}) \right] \\
			G_{SI}^{\mathrm{eq}}(\boldsymbol{Q}) &\equiv \frac{1}{2K} \sum_l \ln \det \left(\boldsymbol{Q}_l  \right)  - \frac{\beta \alpha \lambda}{2K} \sum_l \sum_{a=1}^n Q_l^{aa} \\
			G_E^{\mathrm{eq}}(\boldsymbol{Q}) &\equiv \ln \int \prod_{l} D_{\boldsymbol{Q}_l}\boldsymbol{\lambda}_l  \, e^{-\beta \sum_{a=1}^n \ell\left(f_\star\left[\frac{1}{\sqrt{K}} \sum_l c_l \, \varphi(\lambda_l^{0})\right], \, \frac{1}{\sqrt{K}} \sum_l c_l \, \varphi(\lambda_l^a) \right)}
		\end{align}
	\end{subequations}
	where we have introduced the notation $D_{\boldsymbol{Q}}\boldsymbol{\lambda}$ to denote the integration over a $n+1$ dimensional Gaussian variable $\boldsymbol{\lambda}$ with vanishing mean vector and covariance matrix $\boldsymbol{Q}$. Note that for notational simplicity we have also rescaled the inverse temperature $\beta \to \alpha \beta$; this will not affect the computation of the free entropy of the system apart from an overall scale.

	\subsection{Large width limit}
	We now perform the limit of large number of hidden units $K\to\infty$ that we take only \emph{after} the limits $P,N\to\infty$. Equivalently, we first determine the saddle points and evaluate the functions $G^{\mathrm{eq}}_{SI}$ and $G^{\mathrm{eq}}_E$ at those saddle point values, and only afterwards we consider their large $K$ limit. Hence, although the number of hidden units is infinite, it remains of smaller order than both $N$ and $P$. The limit can be done by noticing that each hidden unit accesses an uncorrelated portion of the input, so we expect the overlap matrix $Q^{ab}_l$ to be independent of $l$
	\begin{equation}
		Q_l^{ab} = Q^{ab} \,.
	\end{equation}	
	We can exploit this property to substantially simplify both the entropic term $G^{\mathrm{eq}}_{SI}$ and energetic contributions $G^{\mathrm{eq}}_E$. The entropic term is straightforward and reads
	\begin{equation}
		G_{SI}^{\mathrm{eq}}(\boldsymbol{Q}) = \frac{1}{2} \ln \det \left( \boldsymbol{Q}  \right)  - \frac{\beta \alpha \lambda}{2} \sum_{a=1}^n Q^{aa} \,.
	\end{equation}
	The energetic term is considerably more complicated to treat~\cite{baldassi2019properties}; but it can be simplified by noticing that the arguments of the loss function $\frac{1}{\sqrt{K}} \sum_l c_l \, \varphi(\lambda_l^{0})$ and $\frac{1}{\sqrt{K}} \sum_l c_l \, \varphi(\lambda_l^a)$ behave as Gaussian random variables. Following closely~\cite{baldassi2019properties, annesi2025exact} one has
	\begin{equation}
		G_E^{\mathrm{eq}}(\boldsymbol{Q}) = \ln \int D_{\boldsymbol{\mathcal{K}}}\boldsymbol{\lambda}  \, e^{- \beta \sum_{a=1}^n \ell\left(f_\star\left[ \lambda^{0}\right], \, \lambda^a\right)}
	\end{equation}
	where the covariance matrix $\boldsymbol{\mathcal{K}}$ now reads
	\begin{equation}
		\label{eq::nngp_kernel_generic}
		\boldsymbol{\mathcal{K}}_{ab} \equiv \int D_{\boldsymbol{Q}} \boldsymbol{\lambda} \, \varphi(\lambda_a)\varphi(\lambda_b) \,.
	\end{equation}
	We have used the assumptions~\eqref{eq::conditions_c} on the second layer weights. 
	
	Note that~\eqref{eq::nngp_kernel_generic} corresponds to the  Neural Network
	Gaussian Process (NNGP) kernel that appears as the covariance matrix of the function
	implemented by a neural network at initialization in the infinite
	width limit and given two different inputs~\cite{Neal1996, Williams1996, zavatone2022NNKernels}.

	\subsection{RS ansatz} \label{sec::equilibrium_RS}
	In order to perform the extremization and the limit $n\to0$ in~\eqref{eq::equilibrium_before_ansatz} we look for a saddle point matrix $\boldsymbol{Q}$ having a particular structure. The simplest ansatz is the \emph{Replica Symmetric} one, which applied to the blocks of $\boldsymbol{Q}$ as in~\eqref{eq::Q_blocks} reads 
	\begin{subequations}
		\label{eq::equilibrium_RS_ansatz}
		\begin{align}
			q^{ab} &= Q \, \delta_{ab} + (1-\delta_{ab}) \, q \\
			r^a &= r \,.
		\end{align}
	\end{subequations}
	It is now an easy job to perform the $n\to 0$ limit. One obtains
	\begin{subequations}
		\begin{align}
			\phi^{\mathrm{eq}} &= \mathrm{extr}_{q, Q, r} \left[ \mathcal{G}_{SI}^{\mathrm{eq}} + \alpha \mathcal{G}_E^{\mathrm{eq}} \right] \\
			\mathcal{G}_{SI}^{\mathrm{eq}} &\equiv \frac{q-r^2}{2(Q-q)} + \frac{1}{2} \ln\left( Q-q \right) - \frac{\beta \alpha \lambda}{2} Q\\
			\mathcal{G}_E^{\mathrm{eq}} &\equiv \int Dv Dx_1  \ln \int Dx_0  \, e^{ - \beta \ell(y(v), \, \hat y(v,x_0, x_1)) }
		\end{align}
	\end{subequations}
	where $y(v)$ and $\hat y(v, x_0, x_1)$ are defined as
	\begin{subequations}
		\label{eq::y_yhat_large_width}
		\begin{align}
			y(v) &\equiv f_\star\left(\sqrt{\Phi_\star}v\right) \,,\\
			\hat y(v, x_0, x_1) &\equiv \frac{D_0}{\sqrt{\Phi_\star}} v + \sqrt{\Phi(Q) - \Phi(q)} \, x_0 + \sqrt{\Phi(q) - \Phi(0) - \frac{D_0^2}{\Phi_\star}} \, x_1 \,,
		\end{align}
	\end{subequations}
	The quantities $\Phi_\star$, $D_0$, $\Phi(\bullet)$ can be compactly expressed in terms of the NNGP kernel function
	\begin{equation}
		\label{eq::nngp_kernel}
		\mathcal{K}(Q_\star, Q, q) \equiv \int Dx Dy \, \varphi(\sqrt{Q_\star} x) \, \varphi\left( \frac{q}{\sqrt{Q_\star}} x + \sqrt{Q - \frac{q^2}{Q_\star}} y \right)
	\end{equation}
	as
	\begin{subequations}
		\label{eq::equilibrium_RS_effective_parameters}
		\begin{align}
			\Phi_\star &= \mathcal{K}(1,1,1) - \mathcal{K}(1,1,0) 
			\\
			D_0 &= \mathcal{K}(1,Q,r) - \mathcal{K}(1,Q,0) \\
			\Phi(q) &= \mathcal{K}(Q, Q, q) 
		\end{align}
	\end{subequations}

	\subsection{$k$-steps Replica Symmetry Breaking} \label{sec::kRSB}
	The equilibrium free entropy in~\eqref{eq::equilibrium_before_ansatz} can be obtained by searching for an extremizer in a more complex set of overlap matrices, which require the breaking of the replica symmetry~\cite{Parisi1980}. We employ here a k-step Replica Symmetry Breaking ($k$RSB) ansatz for the overlap matrix $q^{ab}$
	\begin{equation} 
		\label{eq::1RSBansatz}
		q^{ab} = q_0 + \sum_{s=0}^k (q_{s+1}-q_s) I_{ab}^{(n,m_s)}
	\end{equation}
	where $I_{ab}^{(n,m)}$ is the $(a,b)$ element of a block matrix of size $n \times n$ whose diagonal blocks have size $m \times m$; those  $m\times m$ blocks are filled with ones whereas outside them all the elements are zero. In the previous equation we have also used the definitions
	\begin{equation}
		m_k = 1; \qquad q_{k+1} = Q\,.
	\end{equation}
	For the teacher-student overlap $r^a$ we keep the same ansatz as in~\eqref{eq::equilibrium_RS_ansatz}. 
	We remind here that the $k$-steps RSB matrix $q^{ab}$ has $k+2$ eigenvalues~\cite{annesi2025exact} that read
	\begin{equation}
		\lambda_r \equiv \sum_{s=r}^{k} (q_{s+1} - q_s) m_s \,,  \qquad d_r = n\left(\frac{1}{m_r} - \frac{1}{m_{r-1}}\right), \qquad r = -1 \,, \dots \,, k
	\end{equation}
	having defined $m_{-1} = n \to 0$ and $q_{-1} = 0$, $m_{-2} = \infty$. 
	Using the Schur complement formula
	\begin{equation}
		\begin{split}
			\det \boldsymbol{Q} &= \left(1 - r^2 \, \mathbb{I}^T \boldsymbol{q}^{-1} \mathbb{I} \right) \det \boldsymbol{q} = \left(1 - r^2 \, \sum_{ab} q_{ab}^{-1} \right) \det \boldsymbol{q} = \left(1 - r^2 \, n \sum_{s=0}^{k} (q^{-1}_{s+1} - q^{-1}_s) m_s  \right) \det \boldsymbol{q} \\
			&= \left(1 - n \frac{r^2}{\lambda_0}  \right) \det \boldsymbol{q}
		\end{split}
	\end{equation}
	having used the fact that the set of $k$-step RSB matrices form a group. Using results exposed in~\cite{annesi2025exact} the entropic term now reads
	\begin{equation}
		\mathcal{G}_{SI} = \frac{1}{2} \ln(Q - q_k) + \frac{q_0 - r^2}{2\lambda_0} + \sum_{s=0}^{k-1} \frac{1}{2m_s} \ln \left(\frac{\lambda_s}{\lambda_{s+1}}\right)- \frac{\beta \alpha \lambda}{2} Q  
	\end{equation}
	The energetic term instead is
	\begin{equation}
		\begin{split}
			\mathcal{G}_E^{\mathrm{eq}} &\equiv \int Dv Dh \, f_0\left(v, \sqrt{\Phi(q_0) - \Phi(0) -  \frac{D_0^2}{\Phi_\star}}h + \frac{D_0}{\sqrt{\Phi_\star}}v \right) 
		\end{split}
	\end{equation}
	where $f_0$ is defined via the following recursion
	\begin{subequations}
		\begin{align}
			\label{eq::kRSB_recursion_initial_condition}
			f_k(v, h) &= \ln \int Dz \, e^{-\beta \ell\left(f_\star(\sqrt{\Phi_\star}v), \, h + \sqrt{\Phi(Q) - \Phi(q_k)} \, z \right)} \\
			f_s(v, h) &= \frac{1}{m_s} \ln \int Dz \, e^{m_s f_{s+1}\left(v, \, h + \sqrt{\Phi(q_{s+1}) - \Phi(q_s)} \, z \right)} \,, \qquad s = k-1, \dots, 0
		\end{align}
	\end{subequations}
	Those expressions when evaluated to $k=0, 1$ coincide with the RS (derived in section~\ref{sec::equilibrium_RS}) and 1RSB approximations. 
	
	Finally we point out that the equilibrium free entropy $\phi^{\rm eq}$ needs to be optimized with respect to the overlaps $q_s$, $s=0, \dots, k$ and the breaking-points $m_s$, $s=0, \dots, k-1$ as well as the squared norm $Q$ and the teacher-student overlap $r$.

	\subsubsection{Large $\beta$ limit}
	In the low temperature limit $\beta \to \infty$ the order parameter $q_k$ and the breaking points $m_s$ scale with $\beta$ as
	\begin{subequations}
		\label{eq::kRSB_scaling_large_beta}
		\begin{align}
			q_k &= Q - \frac{\delta q}{\beta} \\
			m_s &= \frac{\delta m_s}{\beta} \,, \qquad s = 0, \dots, k-1 
		\end{align}
	\end{subequations}  
	This implies the following scaling in $\beta$ for $\Phi_0$
	\begin{subequations}
		\begin{align}
			\Phi(Q) - \Phi(q_k) &= 
			\frac{\delta \Phi}{\beta}
		\end{align}
	\end{subequations}
	$\delta \Phi$ can be written as
	\begin{equation}
		\delta \Phi  = \mathcal{K}^{(1,1)}(Q,Q,Q) \, \delta q
	\end{equation}
	in terms of a generalized NNGP kernel function
	\begin{equation}
		\label{eq::generalized_nngp_kernel}
		\mathcal{K}^{(l, m)}(Q_\star, Q, q) \equiv \int Dx Dy \, \varphi^{(l)}(\sqrt{Q_\star} x) \varphi^{(m)}\left( \frac{q}{\sqrt{Q_\star}} x + \sqrt{Q - \frac{q^2}{Q_\star}} y \right)
	\end{equation}
	where $\varphi^{(l)}$ denotes the $l$-th derivative of the activation function $\varphi$. Using this scaling the free energy $ f_\beta = - \lim_{N \to \infty}\frac{1}{\beta N} \mathbb{E}_{\mathcal{D}}\ln Z_\beta$ is finite and can be written as
	\begin{equation}
		- f_{\infty} = \lim_{\beta \to \infty} \frac{\mathcal{G}^{\rm eq}_{SI}}{\beta} + \alpha \lim_{\beta \to \infty} \frac{\mathcal{G}^{\rm eq}_E}{\beta}
	\end{equation}
	The entropic part reads in the limit as
	\begin{equation}
		\lim\limits_{\beta \to \infty} \frac{\mathcal{G}_{SI}}{\beta} = \frac{q_0 - r^2}{2 \delta \lambda_0} + \sum_{s=0}^{k-1} \frac{1}{2\delta m_s} \ln \left(\frac{ \delta \lambda_s}{\delta \lambda_{s+1}}\right)- \frac{\alpha \lambda}{2} Q  
	\end{equation}
	where we have defined
	\begin{subequations}
		\begin{align}
			\delta\lambda_s &= \delta q + (Q - q_{k-1}) \delta m_{k-1} + \sum_{r=s}^{k-2} (q_{r+1} - q_r) \delta m_r \,, \qquad s = 0\,, \dots\,,k-1 \\
			\delta\lambda_k &= \delta q 
		\end{align}
	\end{subequations}
	The energetic term instead reads
	\begin{equation}
		\begin{split}
			\lim_{\beta \to \infty} \frac{\mathcal{G}_E^{\mathrm{eq}}}{\beta} &= \int Dv Dh \, f_0\left(v, \sqrt{\Phi(q_0) - \Phi(0) -  \frac{D_0^2}{\Phi_\star}}h + \frac{D_0}{\sqrt{\Phi_\star}}v \right) 
		\end{split}
	\end{equation}
	where now the $f_0$ satisfies the modified following recursion
	\begin{subequations}
		\label{eq::kRSB_Ge_recursion_large_beta}
		\begin{align}
			f_k(v, h) &= \max_{z}\left[ - \frac{z^2}{2} - \ell(f_\star(\sqrt{\Phi_\star} v), h + \sqrt{\delta \Phi } \, z)\right] \\
			f_s(v, h) &= \frac{1}{\delta m_s} \ln \int Dz \, e^{ \delta m_s f_{s+1}\left(v, \, h + \sqrt{\Phi(q_{s+1}) - \Phi(q_s)} \, z \right)} \,, \qquad s = k-1, \dots, 0
		\end{align}
	\end{subequations}
	
	\subsubsection{Square Loss case}
	In the square loss case the previous recursion present in the energetic term can be fully solved analytically for generic $\beta$ as all the integrals are Gaussian. Indeed one can guess the quantity
	\begin{equation}
		\label{eq::ansatz_recursion_sql}
		f_s(v, h) = A_s - \frac{B_s}{2} \left(h - f_\star\left( \sqrt{\Phi_\star} v \right) \right)^2 \,, \qquad s = 0\,, \dots \,, k
	\end{equation}
	is stable under recursion. The constants $A_s$ and $B_s$ can be derived accordingly. We start determining them by imposing the initial condition $s=k$ as in~\eqref{eq::kRSB_recursion_initial_condition}; one finds
	\begin{subequations}
		\begin{align}
			f_k(v, h) &= -\frac{1}{2} \ln \left( 1 + \beta (\Phi(Q) - \Phi(q_k))  \right) - \frac{\beta (h - f_\star\left( \sqrt{\Phi_\star} v \right))^2}{2(1 + \beta (\Phi(Q) - \Phi(q_k)))}
		\end{align}
	\end{subequations}
	so that 
	\begin{subequations}
		\begin{align}
			A_k &= -\frac{1}{2} \ln \left( 1 + \beta (\Phi(Q) - \Phi(q_k))  \right)  \\
			B_k &= \frac{\beta }{1 + \beta (\Phi(Q) - \Phi(q_k))}
		\end{align}
	\end{subequations}
	If we assume~\eqref{eq::ansatz_recursion_sql} for $f_{s+1}$ one finds for $f_s$
	\begin{equation}
		\begin{split}
			f_s(v, h) &= \frac{1}{m_s} \ln \int Dz \, e^{m_s A_{s+1} - \frac{m_s B_{s+1}}{2} \left( h + \sqrt{\Phi(q_{s+1}) - \Phi(q_s)} \, z  - f_\star\left( \sqrt{\Phi_\star} v \right) \right)^2} \\
			&= A_{s+1} - \frac{1}{2m_s} \ln \left( 1 + m_s B_{s+1} \left( \Phi(q_{s+1}) - \Phi(q_{s}) \right) \right) - \frac{B_{s+1} \left(h - f_\star\left( \sqrt{\Phi_\star} v \right) \right)^2}{2 \left( 1 + m_s B_{s+1} \left( \Phi(q_{s+1}) - \Phi(q_{s}) \right) \right)}
		\end{split}
	\end{equation}
	therefore $A_s$ and $B_s$ satisfy the recursion
	\begin{subequations}
		\begin{align}
			A_s &= A_{s+1} - \frac{1}{2m_s} \ln \left( 1 + m_s B_{s+1} \left( \Phi(q_{s+1}) - \Phi(q_{s}) \right) \right) \\
			B_s &= \frac{B_{s+1}}{1 + m_s B_{s+1} \left( \Phi(q_{s+1}) - \Phi(q_{s}) \right)} = \frac{1}{\frac{1}{B_{s+1}} + m_s \left( \Phi(q_{s+1}) - \Phi(q_{s}) \right)}
		\end{align}
	\end{subequations}
	Those recursions can be solved. Let's start from $B$:
	\begin{equation}
		\begin{split}
			B_s &= \frac{1}{\frac{1}{B_{s+1}} + m_s \left( \Phi(q_{s+1}) - \Phi(q_{s}) \right)} \\
			&= \frac{1}{\frac{1}{B_{s+2}} + m_{s+1} \left( \Phi(q_{s+2}) - \Phi(q_{s+1}) \right) + m_s \left( \Phi(q_{s+1}) - \Phi(q_{s}) \right)} \\ 
			&\;\;\vdots \\
			&= \frac{\beta}{1 + \beta \chi_s}\,, \qquad s = 0\,, \dots\,,k
		\end{split}
	\end{equation}
	having defined 
	\begin{equation}
		\chi_s = \sum_{r=s}^k \left( \Phi(q_{r+1}) - \Phi(q_{r}) \right)  m_r \,, \qquad s = 0\,, \dots\,,k
	\end{equation}
	Correspondingly the coefficients $A_s$ are given by
	\begin{equation}
		\begin{split}
			A_s &= A_{s+1} - \frac{1}{2m_s} \ln \left( 1 + \frac{\beta m_s}{1+\beta \chi_{s+1}} \left( \Phi(q_{s+1}) - \Phi(q_{s}) \right) \right) = A_{s+1} - \frac{1}{2m_s} \ln \left( \frac{1+ \beta \chi_{s}}{1+\beta \chi_{s+1}} \right)\\
			&= - \sum_{r = s}^{k} \frac{1}{2m_r} \ln \left( \frac{1+ \beta \chi_{r}}{1+\beta \chi_{r+1}} \right) = - \sum_{r = s}^{k-1} \frac{1}{2m_r} \ln \left( \frac{1+ \beta \chi_{r}}{1+\beta \chi_{r+1}} \right) - \frac{1}{2} \ln(1+\beta \chi_k) \,, \qquad s=0, \dots\,, k
		\end{split}
	\end{equation}
	Therefore the energetic term reads
	\begin{equation}
		\begin{split}
			\mathcal{G}_E^{\mathrm{eq}} &\equiv \int Dv Dh \, f_0\left(v, \sqrt{\Phi(q_0) - \Phi(0) -  \frac{D_0^2}{\Phi_\star}}h + \frac{D_0}{\sqrt{\Phi_\star}}v \right) \\
			&=  A_0 - \frac{B_0}{2}  \int Dv Dh \,  \left(\sqrt{\Phi(q_0) - \Phi(0) -  \frac{D_0^2}{\Phi_\star}}h + \frac{D_0}{\sqrt{\Phi_\star}}v - f_\star\left( \sqrt{\Phi_\star} v \right)\right)^2
		\end{split}
	\end{equation}
	The remaining integrals over $v$ and $h$ can be solved as well, but the result depends on the task i.e. on $f_\star$. For the classification case one finds
	\begin{equation}
		\begin{split}
			\mathcal{G}_E^{\mathrm{eq}} &=  A_0 - \frac{B_0}{2}  \int Dv Dh \,  \left(\sqrt{\Phi(q_0) - \Phi(0) -  \frac{D_0^2}{\Phi_\star}}h + \frac{D_0}{\sqrt{\Phi_\star}}v - \mathrm{sign}(v)\right)^2 \\
			&= A_0 - \frac{B_0}{2} \left[ \Phi(q_0) - \Phi(0) -  \frac{D_0^2}{\Phi_\star} + \frac{D_0^2}{\Phi_\star} + 1 -2 \frac{D_0}{\sqrt{\Phi_\star}} \int Dv \, v \, \mathrm{sign}(v)  \right]\\
			&=  A_0 - \frac{B_0}{2}  \left[ 1 + \Phi(q_0) - \Phi(0) - 2\sqrt{\frac{2}{\pi}}  \frac{D_0}{\sqrt{\Phi_\star}} \right]
		\end{split}
	\end{equation}
	or
	\begin{equation}
		\mathcal{G}_E^{\mathrm{eq}} = - \frac{\beta}{2(1+\beta \chi_0)} \left(1 + \Phi(q_0) - \Phi(0) - 2 D_0 \sqrt{\frac{2}{\pi \Phi_\star}} \right) - \frac{1}{2} \ln (1 + \beta \chi_k) -  \sum_{s=0}^{k-1} \frac{1}{2m_s} \ln \left( \frac{1 + \beta \chi_s}{1 + \beta \chi_{s+1}}\right) \,,
	\end{equation}
	In the regression instead
	\begin{equation}
		\begin{split}
			\mathcal{G}_E^{\mathrm{eq}} &=  A_0 - \frac{B_0}{2}  \int Dv Dh \,  \left(\sqrt{\Phi(q_0) - \Phi(0) -  \frac{D_0^2}{\Phi_\star}}h + \left(\frac{D_0}{\sqrt{\Phi_\star}} - \sqrt{\Phi_\star}\right)v \right)^2 \\
			&=  A_0 - \frac{B_0}{2}  \left[ \Phi(q_0) - \Phi(0) + \Phi_\star - 2D_0 \right] \\
			&= - \frac{\beta}{2(1+\beta \chi_0)} \left( \Phi(q_0) - \Phi(0) + \Phi_\star - 2D_0 \right) - \frac{1}{2} \ln (1 + \beta \chi_k) -  \sum_{s=0}^{k-1} \frac{1}{2m_s} \ln \left( \frac{1 + \beta \chi_s}{1 + \beta \chi_{s+1}}\right)
		\end{split}
	\end{equation}
	\begin{figure}[t]
		\centering
		\includegraphics[width=0.47\columnwidth]{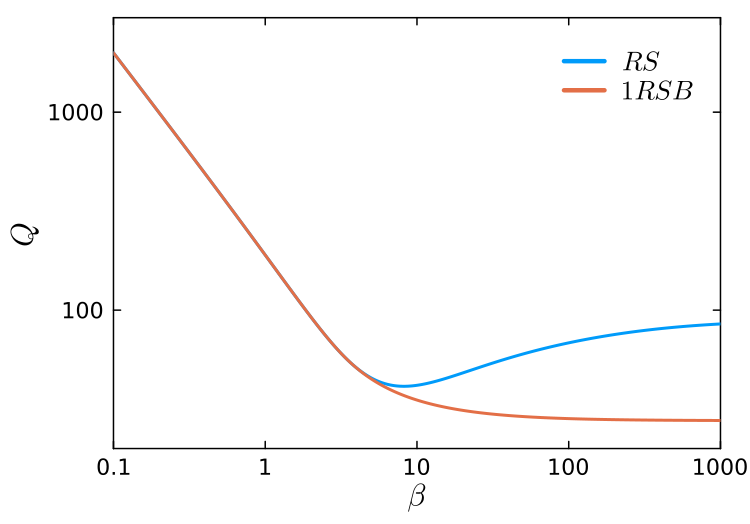}%
		\hfill
		\includegraphics[width=0.47\columnwidth]{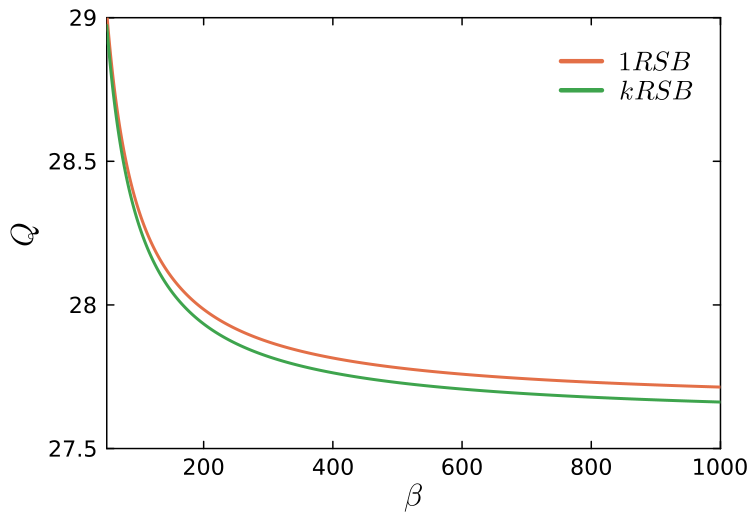}
		\caption{Classification case with square loss function. Dependence of the squared norm $Q$ on the inverse temperature $\beta$ for $\lambda = 0.001$ and $\alpha = 5$.
			Left panel: comparison between the replica symmetric solution and the 1RSB solution.
			Right panel: comparison between the 1RSB solution and the finite-step $k$-RSB approximation with $k=20$. In the $k$-RSB solution we have optimized also on the breaking points $m_s$, $s = 0, \dots, k-1$. In both cases the curves coincide at high temperature and separate as $\beta$ increases.}
		\label{fig:ts_class_difference_1RSB_20RSB_square_norm}
	\end{figure}
	Because the free entropy is analytic in terms of $q_s$, $s=0, \dots, k$, $m_s$, $s=0, \dots, k-1$ and $Q$, $r$ it is easy to optimize it for generic $k$. 
	In Figure~\ref{fig:ts_class_difference_1RSB_20RSB_square_norm} we show the result of this optimization in the classification case, by comparing $k=0$ to the cases $k=1$ and $k=20$. In particular, we show that the squared norm $Q$ changes substantially when going from the $k=0$ RS solution to the $k=1$ RSB solution at large $\beta$, while further RSB steps produce only comparatively small corrections to the 1RSB value.
	
	\begin{figure}[H]
		\centering
		\includegraphics[width=0.47\columnwidth]{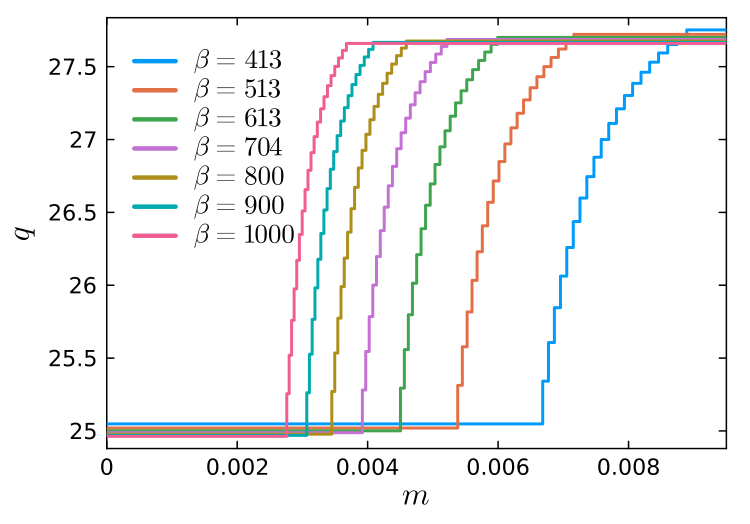} 
		\hfill 
		\includegraphics[width=0.47\columnwidth]{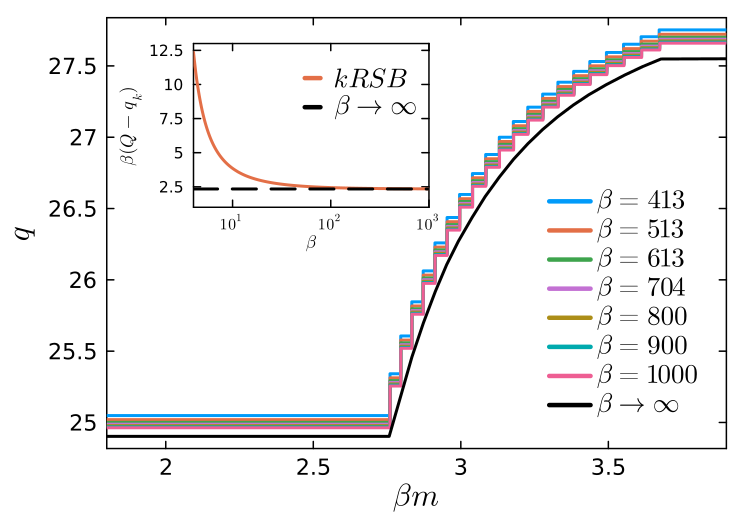}
		\caption{Classification case with square loss function. Overlap function $q_s$ vs breaking points $m_s$ for $\alpha = 5$, $\lambda = 0.001$ and several values of $\beta$ (left panel). On the right panel we have plotted the same data of the left panel vs $\beta m_s \equiv \delta m_s$. As predicted in~\eqref{eq::kRSB_scaling_large_beta} this is the right scaling of the breaking-points for large $\beta$. In black we also depict the exact solution in the limit $\beta \to \infty$. In the inset we show a numerical verification of ~\eqref{eq::kRSB_scaling_large_beta}, plotting $\beta (Q - q_k)$ vs $\beta$ and showing that it approaches a finite limit $\delta q$ (black dashed line).}
		\label{fig:q_finite_beta}
	\end{figure}

	In the left panel of Figure~\ref{fig:q_finite_beta} we instead show the overlap $q$ vs the breaking point value $m$. In the right panel we show a validation of the large beta scalings introduced in~\eqref{eq::kRSB_scaling_large_beta}. 
	
	The analytical large $\beta$ limit expression of the free entropy can be obtained by applying the scalings in~\eqref{eq::kRSB_scaling_large_beta} or using the same approach above to solve the recursion in~\eqref{eq::kRSB_Ge_recursion_large_beta}. One finds for classification the following expression
	\begin{equation}
		\lim\limits_{\beta \to \infty} \frac{\mathcal{G}_E^{\mathrm{eq}}}{\beta} = - \frac{1}{2(1+ \delta \chi_0)} \left(1 + \Phi(q_0) - \Phi(0) - 2 D_0 \sqrt{\frac{2}{\pi \Phi_\star}} \right) -  \sum_{s=0}^{k-1} \frac{1}{2\delta m_s} \ln \left( \frac{1 + \delta \chi_s}{1 + \delta \chi_{s+1}}\right) \,,
	\end{equation}
	while for regression 
	\begin{equation}
		\lim\limits_{\beta \to \infty} \frac{\mathcal{G}_E^{\mathrm{eq}}}{\beta} = - \frac{1}{2(1+  \delta\chi_0)} \left( \Phi(q_0) - \Phi(0) + \Phi_\star - 2D_0 \right) -  \sum_{s=0}^{k-1} \frac{1}{2 \delta m_s} \ln \left( \frac{1 + \delta \chi_s}{1 + \delta \chi_{s+1}}\right) \,.
	\end{equation}
	In the previous expressions we have defined
	
	\begin{align}
		\delta \chi_s &= \delta \Phi + \left( \Phi(Q) - \Phi(q_{k-1}) \right) \delta m_{k-1} +  \sum_{r=s}^{k-2} \left( \Phi(q_{r+1}) - \Phi(q_{r}) \right) \delta m_r \,, \qquad s = 0\,, \dots\,,k-1 \\
		\delta \chi_k &= \delta \Phi \,.
	\end{align}
	
\begin{figure}[H]
		\centering
		\includegraphics[width=0.47\columnwidth]{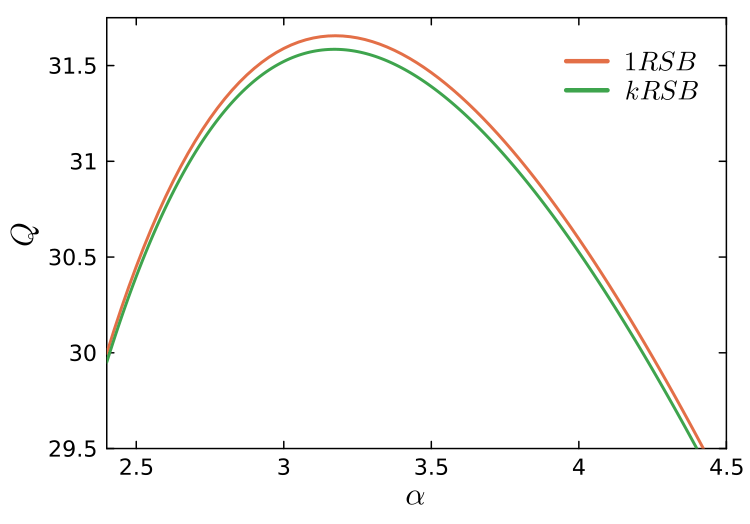} 
		\hfill
		\includegraphics[width=0.47\columnwidth]{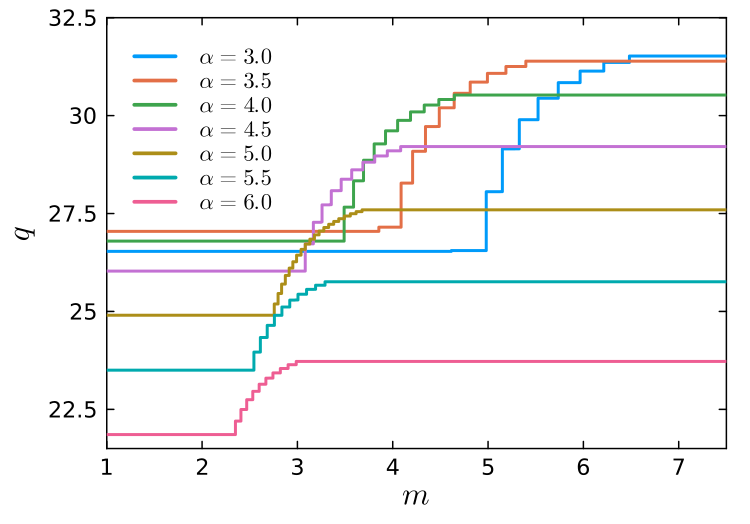}
		\caption{Classification case with square loss function. Left panel: squared norm $Q$ versus $\alpha$ in the large $\beta$ regime, for $\lambda = 0.001$. We compare the 1RSB (orange curve) and $k$-RSB with $k=20$ (green). In the $k$-RSB solution we have optimized also on the breaking points $\delta m_s$, $s = 0, \dots, k-1$; we therefore expect the green curve to be a good proxy of the full-RSB line (i.e. $k\to \infty$). More than one step of RSB are therefore only slightly influencing the squared norm. We point out here that other order parameters are for all purposes left unchanged. Right panel: overlap vs breaking-points for several values of $\alpha$. The maximum value achieved by the overlap is $Q$. }
		\label{fig:ts_class_difference_1RSB_20RSB}
	\end{figure}

	In Figure~\ref{fig:ts_class_difference_1RSB_20RSB} we show the squared norm $Q$ as a function of $\alpha$ in the classification case comparing $k=1$ and $k=20$ steps of RSB in the large $\beta$ limit. We also present some representative plots for the  overlap $q$ in the right panel.

	\subsection{Generalization and training error}
	
	In classification tasks ($f_\star = \mathrm{sign}$) the standard measure of predictive performance after training is the generalization error. It is defined by computing the probability of wrongly predicting a label $y_\star$ which has been generated by the teacher network by sampling a new Gaussian input $\boldsymbol{x}_\star$ (i.e. not in the training set), i.e.
	\begin{equation}
		\epsilon_g = \mathbb{E}_{\boldsymbol{x}_\star} \mathbb{E}_{\mathcal{D}} \left\langle \Theta\left( - y_\star \hat{y}_\star(\boldsymbol{w}) \right) \right\rangle_{\boldsymbol{w} | \mathcal{D}}
	\end{equation}
	where we remind that $\langle \bullet \rangle_{\boldsymbol{w} | \mathcal{D}}$ denotes the average over the Gibbs measure~\eqref{eq::Gibbs_measure} and $\mathbb{E}_{\mathcal{D}}$ is the average over the realization of the dataset. $\hat{y}_\star(\boldsymbol{w})$ denotes the preactivation of the student's output corresponding to the test input $\boldsymbol{x}_\star$. 
	This quantity can be easily computed again by using the replica method. In the $K\to \infty$ limit, in particular, one finds
	\begin{equation}
		\epsilon_g = \frac{1}{\pi} \arccos\left( \frac{D_0}{\sqrt{\Phi_\star (\Phi(Q) - \Phi(0))}} \right)
	\end{equation}
	In regression we instead measure the test loss. In the $K \to \infty$ limit one has
	\begin{equation}
		\epsilon_g = \mathbb{E}_{\boldsymbol{x}_\star} \mathbb{E}_{\mathcal{D}} \frac{1}{2} \left\langle \left( y_\star - \hat{y}_\star(\boldsymbol{w}) \right)^2 \right\rangle_{\boldsymbol{w} | \mathcal{D}} = \frac{\Phi(Q) - \Phi(0) + \Phi_\star - 2D_0}{2}
	\end{equation}
	The training loss and the training error respectively relevant in regression and classification tasks can be found similarly. For example the training error is given by the following expression
	\begin{equation}
		\epsilon_t = \mathbb{E}_{\mathcal{D}} \left\langle \frac{1}{\alpha N} \sum_{\mu}\Theta\left( - y \hat{y}(\boldsymbol{w}) \right) \right\rangle_{\!\boldsymbol{w} | \mathcal{D}} = \left\langle \Theta\left(- y(v)  \, \hat y(v,x_0, x_1, x_{01})\right) \right\rangle_\beta\,.
	\end{equation}

	\section{Replica computation of the spectrum of the Hessian} \label{sec::Hessian_computation}
	
	\subsection{Gibbs formulation of the spectrum of the Hessian} \label{app::Gibbs_Hessian}
	Consider $\mathcal{H}$ to be a generic random matrix of size $N\times N$. Its limiting spectral distribution is defined as
	\begin{equation}
		\rho_\mathcal{H}(z) = \lim_{N\to\infty}\frac{1}{N} \sum_{i=1}^{N} \delta(z - \lambda_i)
	\end{equation}
	where we have denoted by $\lambda_i$, $i \in [N]$ the eigenvalues of $\mathcal{H}$. A fundamental quantity related to the empirical spectral distribution $\rho_\mathcal{H}(z)$ is the so-called Stieltjes transform, which is the normalized trace of the resolvent of $\mathcal{H}$
	\begin{equation}
		\mathcal{R}_\mathcal{H}(z) \equiv  \lim_{N\to\infty} \frac{1}{N} \text{Tr} \left(z I - \mathcal{H}\right)^{-1} =  \lim_{N\to\infty}\frac{1}{N} \sum_{i=1}^{N} \frac{1}{z - \lambda_i} = \int d\lambda \, \frac{\rho_\mathcal{H}(\lambda)}{z - \lambda}	
	\end{equation}
	The previous expression can be inverted as
	\begin{equation}
		\label{eq::spectrum}
		\rho_\mathcal{H}(z) = \lim\limits_{\varepsilon \to 0^+} \frac{1}{\pi} \text{Im} \mathcal{R}_\mathcal{H}(z - i \varepsilon)
	\end{equation}
	The Stieltjes transform can be computed by using the replica method, since it can be written as the derivative of a quantity that plays the role of a ''free entropy'' $\phi_\mathcal{H}(z)$
	\begin{equation}
		\mathcal{R}_\mathcal{H}(z) = \frac{\partial}{\partial z} \phi_\mathcal{H}(z)
	\end{equation}
	which is defined as
	\begin{subequations}
		\begin{align}
			\phi_\mathcal{H}(z) &= -\frac{2}{N} \ln \mathcal{Z}_\mathcal{H}(z) \\
			\mathcal{Z}_\mathcal{H}(z) &= \det(\mathcal{H} - zI)^{-1/2} = \int \prod_{i=1}^{N} \frac{du_i}{\sqrt{2\pi}} \, e^{- \frac{1}{2} \sum_{ij} u_i (\mathcal{H}_{ij} - z \delta_{ij}) u_j }
			\label{eq::partition_function_Hessian}
		\end{align}
	\end{subequations}
	$\mathcal{Z}_\mathcal{H}(z)$ plays the role of a partition function. 
	Notice that similarly the cumulative distribution of $\mathcal{H}$ can be obtained by computing the imaginary part of $\phi_\mathcal{H}$
	\begin{equation}
		\mathcal{C}_\mathcal{H}(x) = \int_{-\infty}^{x} dz \, \rho_\mathcal{H}(z) = \frac{1}{\pi} \text{Im} \phi_\mathcal{H}(x) \,.
	\end{equation}
	The previous identities are valid for a generic $N \times N$ random matrix $\mathcal{H}$. 
	In the following we will be interested in computing the spectrum of the Hessian of the loss function of the teacher-student tree committee machine model as defined in section~\ref{sec::setting&objective}, for a typical weight sampled from the Gibbs measure~\eqref{eq::Gibbs_measure}. This means that we need to compute
	\begin{equation}
		- 2\phi(z) \equiv \lim\limits_{N\to \infty} \mathbb{E}_{\mathcal{D}}\langle \phi_\mathcal{H}(z) \rangle_{\boldsymbol{w}|\mathcal{D}} = - \lim\limits_{N\to \infty} \frac{2}{N} \mathbb{E}_{\mathcal{D}} \langle  \ln \mathcal{Z}_\mathcal{H}(z) \rangle_{\boldsymbol{w} | \mathcal{D}}
	\end{equation}
	where $\langle \bullet \rangle_{\boldsymbol{w} | \mathcal{D}}$ denotes the average over the Gibbs measure~\eqref{eq::Gibbs_measure} and $\mathbb{E}_{\mathcal{D}}$ is the average over the realization of the dataset. 
	
	The Hessian
	\begin{equation}
		\alpha \mathcal{H} 
		=
		\sum_{\mu=1}^{P}  \frac{\partial^2 \ell\!\left(y^\mu,\hat y^\mu(\w)\right)}{\partial \boldsymbol{w} \partial \boldsymbol{w}^T}
	\end{equation}
	is in our case an $N\times N$ matrix having the following block structure
	\begin{equation}
		\alpha \mathcal{H}_{li, l'i'} = \frac{1}{N} \sum_\mu \ell''(y^\mu, \hat y^\mu) c_l c_{l'} \varphi_l' \varphi_{l'}' x_{li}^\mu x_{l'i'}^\mu + \delta_{l l'} \frac{\sqrt{K}}{N}\sum_\mu \ell'(y^\mu, \hat y^\mu) c_l \varphi_l''  x_{li}^\mu x_{li'}^\mu \,.
		\label{eq:HessianCommittee}
	\end{equation}
	where we have used the notation $\varphi_{l} =  \varphi \left( \sqrt{\frac{K}{N}} \sum_{i=1}^{N/K} w_{li} x_{li}^\mu \right)$ and we have denoted by a prime the derivative of the loss function with respect to its second argument $\hat y$.  Notice that we have not included in this definition the second derivative with respect to $\boldsymbol{w}$ of the L2 regularization appearing in the loss $\mathcal{L}(\boldsymbol{w})$ in~\eqref{eq:loss_main}. Since the regularization term is quadratic in $\boldsymbol{w}$ this provides only a shift to the spectrum of the Hessian depending on the value of $\lambda$. Notice that the Hessian that we compute here differs from the one in the main text by a scaling factor $\alpha$; this factor can be taken into account a posteriori by appropriate rescaling of the spectrum: $\rho_\mathcal{H}(z) = \alpha \rho_\mathcal{\alpha H}( \alpha z)$. 
	
	\subsection{Replica tricks}
	
	In order to average over the input data, we use two replica tricks, one for $1/Z_\beta = \lim\limits_{n\to 0} Z_\beta^{n-1}$ and $\ln \mathcal{Z}_{\alpha \mathcal{H}} = \lim\limits_{s\to 0} \partial_s \mathcal{Z}^s_{\alpha \mathcal{H}}$. Notice indeed the similarity of our computation with the standard Franz-Parisi potential~\cite{franz1995recipes}. By analogy the variables $u_{li}$ are acting as ``fake'' weights whereas the variable $z$ of the spectral density $\rho_\mathcal{\alpha H}(z)$ acts like a regularization over those fake weights. 
	\begin{equation}
		\label{eq::phi(z)}
		\phi(z) = \lim\limits_{N\to \infty} \frac{1}{N} \mathbb{E}_{\mathcal{D}} \langle  \ln \mathcal{Z}_{\alpha \mathcal{H}}(z) \rangle_{\boldsymbol{w} | \mathcal{D}} 
		= \lim\limits_{N\to \infty}  \lim\limits_{\substack{n\to 0 \\ s\to0}} \frac{\partial_s \tilde{\phi}}{N}
	\end{equation}
	where
	\begin{equation}
		\label{eq::tildephi(z)}
		\tilde{\phi} \equiv \mathbb{E}_{\mathcal{D}} \, \int \prod_{a=1}^n d \boldsymbol{w}^a \, e^{- \beta \sum_{a=1}^n \mathcal{L}(\boldsymbol{w}^a)} \, \mathcal{Z}^s_{\mathcal{H}}(z)
	\end{equation}
	In the following we will denote by $a,b \in [n]$ whereas $c, d \in [s]$. Using the expression of the Hessian in~\eqref{eq:HessianCommittee}, the quadratic term in the partition function in~\eqref{eq::partition_function_Hessian} is given by
	\begin{multline}
		\alpha \sum_{li, l'i'} u_{li} \mathcal{H}_{li, l'i'} u_{l'i'} = \sum_\mu \ell''(y^\mu, \hat y^\mu) \left( \frac{1}{\sqrt{K}}\sum_l c_l \varphi'_l \, \sqrt{\frac{K}{N}} \sum_i u_{li} x_{li}^\mu \right)^2 \\
		+ \sum_\mu \ell'(y^\mu, \hat y^\mu) \frac{1}{\sqrt{K}} \sum_l c_l  \left( \sqrt{\frac{K}{N}} \sum_i u_{li} x_{li}^\mu \right)^2 \varphi_l'' 
	\end{multline}
	We therefore found that the dependence on the disorder in the right hand side of~\eqref{eq::tildephi(z)} appears via the three variables
	\begin{subequations}
		\begin{align}
			\eta^\mu_l &= \sqrt{\frac{K}{N}} \sum_{i=1}^{N/K} w_{li}^\star x_{li}^\mu \\
			\lambda^\mu_{la} &= \sqrt{\frac{K}{N}} \sum_{i=1}^{N/K} w_{li}^a x_{li}^\mu \\ 
			h_{lc}^{\mu} &=\sqrt{\frac{K}{N}}\sum_{i=1}^{N/K}u_{li}^cx_{li}^{\mu} 
		\end{align}
	\end{subequations}
	which are jointly Gaussian random variables. Equivalently we can enforce those definitions via delta functions insertions and represent those delta functions via their Fourier representation. The average over the input data reads
	\begin{equation}
		\begin{split}
			&\prod_{li} \prod_\mu \mathbb{E}_{x_{li}^\mu} e^{-i\sqrt{\frac{K}{N}} x_{li}^{\mu}\left(\hat{\eta}_{l}^{\mu}w_{li}^\star+\sum_{a}\hat{\lambda}_{la}^{\mu}w_{li}^{a}+\sum_{c}\hat{h}_{lc}^{\mu}u_{li}^{c}\right)} = \prod_{li} \prod_\mu  e^{-\frac{K}{2N} \left(\hat{\eta}_{l}^{\mu}w_{li}^\star+\sum_{a}\hat{\lambda}_{la}^{\mu}w_{li}^{a}+\sum_{c}\hat{h}_{lc}^{\mu}u_{li}^{c}\right)^2}\\
			&= \prod_{\mu l}  e^{- \frac{1}{2}(\hat{\eta}^\mu_{l})^2 - \frac{1}{2} \sum_{ab} q_{l}^{ab} \hat \lambda_{la}^{\mu} \hat \lambda_{lb}^{\mu} - \frac{1}{2} \sum_{cd} p_{l}^{cd} \hat h_{lc}^{\mu} \hat h_{ld}^{\mu} - \sum_{ac} t_{l}^{ac} \hat \lambda^\mu_{la} \hat h^\mu_{lc} - \hat \eta_{l}^\mu \sum_{a} r_l^a \hat \lambda^\mu_{la}  - \hat \eta_{l}^\mu \sum_{c} o_l^c \hat h^\mu_{lc}}
		\end{split}
	\end{equation}
	where we have introduced the following overlap quantities
	\begin{subequations}
		\begin{align}
			q_l^{ab} &\equiv \frac{K}{N} \sum_{i=1}^{N/K} w_{li}^a w_{li}^b\,, \qquad r_l^{a} \equiv \frac{K}{N} \sum_{i=1}^{N/K} w_{li}^\star w_{li}^a \\
			p_l^{cd} &\equiv \frac{K}{N} \sum_{i=1}^{N/K} u_{li}^c u_{li}^d\,,  \qquad t_l^{ac} \equiv \frac{K}{N} \sum_{i=1}^{N/K} w_{li}^a u_{li}^c\,, \qquad o_l^{c} \equiv \frac{K}{N} \sum_{i=1}^{N/K} w_{li}^\star u_{li}^c
		\end{align}
	\end{subequations}
	and we have also used the fact that $\frac{K}{N} \sum_i (w_{li}^\star)^2 = 1$ for large $N$ thanks to the central limit theorem. The previous overlaps can be enforced via delta functions. In the end one finds the free entropy to be
	\begin{equation}
		\tilde{\phi} = \int\prod_{abl}\frac{dq^{ab}_{l} d\hat{q}^{ab}_{l}}{2\pi}\prod_{acl}\frac{dt^{ac}_{l} d\hat{t}^{ac}_{l}}{2\pi}\prod_{cd l}\frac{dp^{cd}_{l} d\hat{p}^{cd}_{l}}{2\pi}\prod_{al}\frac{dr^{a}_{l} d\hat{r}^{a}_{l}}{2\pi} \prod_{cl}\frac{ do^{c}_{l} d\hat{o}^{c}_{l}}{2\pi}e^{ N\left(G_{I}+G_{S}+\alpha G_{E}\right)}
	\end{equation}
	
	where the interaction, entropic and energetic terms are the following:
	\begin{subequations}
		\begin{align}
			G_{I}&=-\frac{1}{2K}\sum_{abl}q^{ab}_{l} \hat{q}^{ab}_{l} - \frac{1}{K} \sum_{acl}t^{ac}_{l} \hat{t}^{ac}_{l}-\frac{1}{2K}\sum_{cdl}\hat{p}^{cd}_{l}p^{cd}_{l} - \frac{1}{K}\sum_{al}r^{a}_{l} \hat{r}^{a}_{l} - \frac{1}{K}\sum_{cl}o^{c}_{l} \hat{o}^{c}_{l} + \frac{z}{2K}\sum_{lc} p^{cc}_{l}\\
			G_{S} &= \frac{1}{K}\log\mathbb{E}_{\boldsymbol{w}^\star} \! \int\prod_{la} dw_l^a \prod_{lic}\frac{du_{li}^{c}}{\sqrt{2\pi}}e^{\frac{1}{2}\sum_{abl}\hat{q}^{ab}_{l} w_{l}^{a}w_{l}^{b}+\frac{1}{2}\sum_{cdl}\hat{p}^{cd}_{l} u_{l}^{c}u_{l}^{d}+\sum_{acl}\hat{t}^{ac}_{l} w_{l}^{a}u_{l}^{c} +\sum_{al}\hat{r}^{a}_{l} w_{l}^\star w_{l}^{a}+\sum_{cl}\hat{o}^{c}_{l} u_{l}^{c} w_{l}^\star}\\
			G_{E}	&=\log \int \prod_{l}\frac{ d\eta_{l} d\hat{\eta}_{l}}{2\pi}\prod_{al}\frac{d\lambda_{la} d\hat{\lambda}_{la}}{2\pi}\prod_{cl}\frac{d\hat{h}_{lc} dh_{lc}}{2\pi}  e^{i\sum_{l}\eta_{l} \hat{\eta}_{l} +i\sum_{la}\lambda_{la} \hat{\lambda}_{la} +i\sum_{lc}h_{lc} \hat{h}_{lc}} \\
			&\nonumber \times e^{-\beta\sum_{a}\ell\left(y(\boldsymbol{\eta}),\, \hat{y}(\boldsymbol{\lambda}_{a})\right)} \\ 
			&\nonumber \times e^{-\frac{1}{2}\sum_{l}\left(\hat{\eta}_{l}^2 + \sum_{ab}q^{ab}_{l} \hat{\lambda}_{la}\hat{\lambda}_{lb} + \sum_{cd} p^{cd}_{l} \hat{h}_{lc} \hat{h}_{ld} + 2\sum_{ac} t^{ac}_{l}\hat{\lambda}_{la} \hat{h}_{lc}  + 2 \hat{\eta}_{l} \sum_{a} r^{a}_{l} \hat{\lambda}_{la} + 2\hat{\eta}_{l} \sum_{c} o^{c}_{l} \hat{h}_{lc} \right)}\\
			&\nonumber \times e^{-\frac{1}{2}  \ell''(y(\boldsymbol{\eta}), \, \hat y(\boldsymbol{\lambda}_1)) \sum_c \left( \frac{1}{\sqrt{K}}\sum_l c_l \varphi'(\lambda_{l1}) \, h_{lc} \right)^2 -\frac{1}{2} \ell'(y(\boldsymbol{\eta}), \, \hat y(\boldsymbol{\lambda}_1)) \frac{1}{\sqrt{K}} \sum_{lc} c_l h_{lc}^2 \varphi''(\lambda_{l1}) }
		\end{align}
	\end{subequations}
	where we have introduced the notation 
	\begin{subequations}
		\begin{align}
			y(\boldsymbol{\eta}) &= f_\star \left[ \frac{1}{\sqrt{K}} \sum_l c_l \, \varphi \!\left( \eta_l \right) \right] \\
			\hat y^\mu(\boldsymbol{\lambda}_a) &= \frac{1}{\sqrt{K}} \sum_l c_l \, \varphi \left(\lambda_{la} \right)
		\end{align}
	\end{subequations}
	Note that the average over $\boldsymbol{w}^\star$ can be performed in the entropic term $G_S$. and we can simplify its expression to
	\begin{equation}
		G_S = \frac{1}{K}\log \int\prod_{la} dw_l^a \prod_{lic}\frac{du_{li}^{c}}{\sqrt{2\pi}}e^{\frac{1}{2}\sum_{abl} (\hat{q}^{ab}_{l} + \hat r^a_l \hat r^b_l) w_{l}^{a}w_{l}^{b}+\frac{1}{2}\sum_{cdl}(\hat{p}^{cd}_{l}+ \hat{o}^{c}_{l} \hat{o}^{d}_{l}) u_{l}^{c}u_{l}^{d}+\sum_{acl}(\hat{t}^{ac}_{l}+ \hat{r}^{a}_{l} \hat{o}^{d}_{l}) w_{l}^{a}u_{l}^{c} }
	\end{equation}
	Similarly it is convenient to perform the integral over $\hat \eta_{l}$ in the energetic term
	\begin{equation}
		\begin{split}
			G_{E}	&=\log \int \prod_{l} D \eta_l \prod_{al}\frac{d\lambda_{la} d\hat{\lambda}_{la}}{2\pi}\prod_{cl}\frac{d\hat{h}_{lc} dh_{lc}}{2\pi}  e^{ i\sum_{la}(\lambda_{la} +r_l^a \eta_l)\hat{\lambda}_{la} +i\sum_{lc} (h_{lc} + o_l^c \eta_l)\hat{h}_{lc} -\beta\sum_{a}\ell\left(y(\boldsymbol{\eta}),\, \hat{y}(\boldsymbol{\lambda}_{a})\right)} \\ 
			& \times e^{-\frac{1}{2}\sum_{l}\left( \sum_{ab} (q^{ab}_{l} - r_l^a r_l^b) \hat{\lambda}_{la}\hat{\lambda}_{lb} + \sum_{cd} (p^{cd}_{l} - o_l^c o_l^d ) \hat{h}_{lc} \hat{h}_{ld} + 2\sum_{ac} (t^{ac}_{l} - r_l^a o_l^c)\hat{\lambda}_{la} \hat{h}_{lc}  \right)}\\
			& \times e^{-\frac{1}{2}  \ell''(y(\boldsymbol{\eta}), \, \hat y(\boldsymbol{\lambda}_1)) \sum_c \left( \frac{1}{\sqrt{K}}\sum_l c_l \varphi'(\lambda_{l1}) \, h_{lc} \right)^2 -\frac{1}{2} \ell'(y(\boldsymbol{\eta}), \, \hat y(\boldsymbol{\lambda}_1)) \frac{1}{\sqrt{K}} \sum_{lc} c_l h_{lc}^2 \varphi''(\lambda_{l1}) }
		\end{split}
	\end{equation}
	where we remind the notation $D\eta \equiv \frac{d\eta}{\sqrt{2\pi}} e^{-\eta^2/2}$.

	\subsection{Annealed ansatz}
	
	In section~\ref{sec::Equilibrium}, we showed that the 1RSB theory provides an accurate approximation of the equilibrium measure, while higher-order RSB effects yield only negligible corrections. For simplicity, we therefore adopt a 1RSB ansatz for the overlap matrix $q_l^{ab}$ (and the usual RS parametrization for $r_l^a$)
	\begin{subequations}
		\begin{align}
			q_l^{ab} &= q_0 + (q_1-q_0) I_{ab}^{(n,m)} + (Q-q_1) I_{ab}^{(n,1)} \\
			\hat q_l^{ab} &= \hat q_0 + (\hat q_1-\hat q_0) I_{ab}^{(n,m)} + (\hat Q-\hat q_1) I_{ab}^{(n,1)} \\
			r_l^{a} &= r \\
			\hat r_l^{a} &= \hat r
		\end{align}
	\end{subequations}
	where we remind that $I_{ab}^{(n,m)}$ is the $(a,b)$ element of a block matrix of size $n \times n$ whose diagonal blocks have size $m \times m$; those  $m\times m$ blocks are filled with ones whereas outside them all the elements are zero. We do not have to worry of those order parameters because they fully satisfy the saddle point equations found in section~\ref{sec::Equilibrium} as in standard Franz-Parisi computations~\cite{franz1995recipes,huang2014origin}. Moreover, the conjugated parameters can be analytically expressed in terms of the non-conjugated ones only. 
	
	We instead consider an annealed ansatz for the new overlaps
	\begin{subequations}
		\label{eq::annealed_ansatz}
		\begin{align}
			p_l^{cd} &= P \delta_{cd}  \\
			\hat p_l^{cd} &= - \hat P \delta_{cd} \\
			t_l^{ac} &= \hat t_l^{ac} = o_l^{c} = \hat o_l^{c} = 0
		\end{align}
	\end{subequations}
	Inserting this ansatz one can easily perform the limit $n\to 0$ and $s\to 0$. The interaction and entropic term read
	\begin{subequations}
		\begin{align}
			\mathcal{G}_{I} &\equiv \lim\limits_{\substack{n\to 0 \\ s\to0}}\partial_s G_{I} =  \frac{1}{2}\hat{P}P + \frac{z}{2}P  \\ 
			\mathcal{G}_{S} &\equiv \lim\limits_{\substack{n\to 0 \\ s\to0}}\partial_s G_{S} = -\frac{1}{2}\ln \hat{P} 
		\end{align}
	\end{subequations}
	The energetic term is more involved, but it can be shown with standard methods to be equal to
	\begin{equation}
		\begin{split}
			\mathcal{G}_E &= \lim\limits_{\substack{n\to 0 \\ s\to0}}\partial_s G_{E}  = \left\langle \log Z_h \right\rangle_\beta \,,
		\end{split}
	\end{equation}
	where $Z_h$ and $\left\langle \bullet \right\rangle_\beta$ which represents the average over the equilibrium measure, are given by
	\begin{subequations}
		\begin{align}
			\left\langle \bullet \right\rangle_\beta = \int D \boldsymbol{v}\,D \boldsymbol{x_{1}} &\;
			\frac{
				\int D\boldsymbol{x_{01}}\;
				\mathcal{N}_0(\boldsymbol{x_1},\boldsymbol{x_{01}},\boldsymbol{v})^{m-1} \int D \boldsymbol{x_0}\;
				e^{
					-\beta\,\ell(
					f_\star(\mathcal{O}_\star(\boldsymbol{v})),\,\mathcal{O}(\boldsymbol{x_0},\boldsymbol{x_1},\boldsymbol{x_{01}},\boldsymbol{v}) )} \; \bullet
			}{
				\int D \boldsymbol{x_{01}}\;
				\mathcal{N}_0(\boldsymbol{x_1},\boldsymbol{x_{01}},\boldsymbol{\eta})^m }\\
			\mathcal{N}_0(\boldsymbol{x_1},\boldsymbol{x_{01}},\boldsymbol{v}) &= \int D \boldsymbol{x_0}\; e^{-\beta\,\ell(
				f_\star(\mathcal{O}_\star(\boldsymbol{v})),\,\mathcal{O}(\boldsymbol{x_0},\boldsymbol{x_1},\boldsymbol{x_{01}},\boldsymbol{v}))} \\
			Z_h(\boldsymbol{x_0},\boldsymbol{x_1},\boldsymbol{x_{01}},\boldsymbol{v})	&= \!
			\int D \boldsymbol{h}\;
			e^{-\frac{P}{2}\,
				\ell''(
				f_\star(\mathcal{O}_\star(\boldsymbol{v})),\,\mathcal{O}(\boldsymbol{x_0},\boldsymbol{x_1},\boldsymbol{x_{01}},\boldsymbol{v}))\,
				\mathcal{O}_1(\boldsymbol{x_0},\boldsymbol{x_1},\boldsymbol{x_{01}},\boldsymbol{v},\boldsymbol{h})^2} \\
			&\nonumber \times e^{- \frac{P}{2}\,
				\ell'(
				f_\star(\mathcal{O}_\star(\boldsymbol{v})),\,\mathcal{O}(\boldsymbol{x_0},\boldsymbol{x_1},\boldsymbol{x_{01}},\boldsymbol{v})
				)\,
				\mathcal{O}_2(\boldsymbol{x_0},\boldsymbol{x_1},\boldsymbol{x_{01}},\boldsymbol{v},\boldsymbol{h})}
		\end{align}
	\end{subequations}
	We have also defined the following quantities
	\begin{subequations}
		\label{eq::O}
		\begin{align}
			\mathcal{O}_\star(\boldsymbol{v})
			&=
			\frac{1}{\sqrt{K}}
			\sum_{l=1}^K c_l\,\varphi(v_l),
			\\
			u_l(x_0,x_1,x_{01},v)
			&=
			\sqrt{Q-q_1}\,x_{0,l}
			+\sqrt{q_1-q_0}\,x_{01,l}
			+\sqrt{q_0-r^2}\,x_{1, l}
			+r\,v_l
			\\
			\mathcal{O}(\boldsymbol{x_0},\boldsymbol{x_1},\boldsymbol{x_{01}},\boldsymbol{v})
			&=
			\frac{1}{\sqrt{K}}
			\sum_{l=1}^K c_l\,\varphi\left(u_l(x_0,x_1,x_{01}, v)\right)
			\\
			\mathcal{O}_1(\boldsymbol{x_0},\boldsymbol{x_1},\boldsymbol{x_{01}},\boldsymbol{v},\boldsymbol{h})
			&=
			\frac{1}{\sqrt{K}}
			\sum_{l=1}^K c_l h_l\,
			\varphi' \left(u_l(x_0,x_1,x_{01},v)\right)
			\\
			\mathcal{O}_2(\boldsymbol{x_0},\boldsymbol{x_1},\boldsymbol{x_{01}},\boldsymbol{v},\boldsymbol{h})
			&=
			\frac{1}{\sqrt{K}}
			\sum_{l=1}^K c_l h_l^2\,
			\varphi''\!\left(u_l(x_0,x_1,x_{01},v)\right) \,.
		\end{align}
	\end{subequations}
	Note that the expression of $Z_h$ can be simplified by using Gaussian integration
	\begin{equation}
		\mathcal{G}_E = - \frac{1}{2} \left \langle \ln \det \left( I_K + P \mathcal{S} \right) \right \rangle_\beta
	\end{equation}
	where we have introduced the $K\times K$ matrix $\mathcal{S}$ with elements
	\begin{equation}
		\mathcal{S}_{l l'} \equiv \frac{\ell''(f_\star(\mathcal{O}_\star), \mathcal{O})}{K} c_l c_{l'} \varphi'(u_l) \varphi'(u_{l'}) + \delta_{l l'} \frac{\ell'(f_\star(\mathcal{O}_\star), \mathcal{O})}{\sqrt{K}} c_l \varphi''(u_l)
	\end{equation}
	Notice also that the saddle point equation for $\hat P$ can be readily solved as $\hat P = 1/P$. The entropic and interaction terms therefore simply read as follows 
	\begin{equation}
		\mathcal{G}_{SI} \equiv \mathcal{G}_S + \mathcal{G}_I 
		= \frac{1}{2} + \frac{1}{2} \ln P +\frac{z}{2}P
	\end{equation}
	We therefore found that the ``free entropy'' $\phi(z)$ in~\eqref{eq::phi(z)} can be found by
	\begin{equation}
		\label{eq::final_phi}
		\phi(z) = \max_{P}\left[ \mathcal{G}_{SI} + \alpha \mathcal{G}_E \right]
	\end{equation}
	The values of $q$, $r$ and $Q$ are found from the equilibrium saddle point equations (see section~\ref{sec::Equilibrium}). 
	The maximization in~\eqref{eq::final_phi} imposes the following saddle point equation for $P$:
	\begin{equation}
		\label{eq::general_stieltjes_transform_equation}
		z = - \frac{1}{P} + \alpha \left\langle \mathrm{Tr} \left[ \mathcal{S} \left(I_K + P \mathcal{S}\right)^{-1} \right] \right\rangle_\beta
	\end{equation}
	Note also that, following the discussion in Appendix~\ref{app::Gibbs_Hessian} the Stieltjes transform of the Hessian is equal to the value of $-P$ with $P$ solving the saddle point equation~\eqref{eq::general_stieltjes_transform_equation}, as
	\begin{equation}
		\mathcal{R}_{\alpha \mathcal{H}}(z) = -2 \frac{\partial}{\partial z} \phi(z) = - P \,.
	\end{equation}
	The spectrum is then computed using~\eqref{eq::spectrum} which gives
	\begin{equation}
		\label{eq::spectrum_P}
		\rho_\mathcal{\alpha H}(z) = - \frac{1}{\pi} \text{Im} P \,.
	\end{equation} 
	
	\subsection{Large width limit}
	
	Before sending $K$ to infinity, we can furthermore simplify the energetic term, by using the matrix determinant lemma
	\begin{equation}
		\begin{split}
			\ln \det \left(I_K + P \mathcal{S}\right) &= \sum_l \log \left(1 + \frac{P \ell'(f_\star(\mathcal{O}_\star), \mathcal{O})}{\sqrt{K}} c_l \varphi''(u_l)  \right) \\
			&+ \log \left[1 + \frac{P \ell''(f_\star(\mathcal{O}_\star), \mathcal{O})}{K} \sum_l \frac{c_l^2 \varphi'(u_l)^2}{1+ \frac{P \ell'(f_\star(\mathcal{O}_\star), \mathcal{O})}{\sqrt{K}} c_l \varphi''(u_l)} \right] \,.
		\end{split}
	\end{equation}
	When $K$ is large we can Taylor expand this expression to get
	\begin{equation}
		\begin{split}
			\ln \det \left(I_K + P \mathcal{S}\right) &\simeq 
			P \ell'(f_\star(\mathcal{O}_\star), \mathcal{O})  \langle \mathcal{O}_2 \rangle_h - \frac{P^2 \ell'(f_\star(\mathcal{O}_\star), \mathcal{O})^2}{4} \left( \langle \mathcal{O}_2^2 \rangle_h - \langle \mathcal{O}_2 \rangle_h^2\right) \\
			&+ \log \left[1 + P \ell''(f_\star(\mathcal{O}_\star), \mathcal{O}) \langle \mathcal{O}_1^2\rangle_h \right] 
		\end{split}
	\end{equation}
	where we have used the identities
	\begin{subequations}
		\begin{align}
			\langle \mathcal{O}_1^2 \rangle_h &= \frac{1}{K} \sum_l c_l^2 \varphi'(u_l)^2 \\
			\langle \mathcal{O}_2 \rangle_h &= \frac{1}{\sqrt{K}} \sum_l c_l \varphi''(u_l) \\
			\langle \mathcal{O}_2^2 \rangle_h - \langle \mathcal{O}_2 \rangle_h^2 &= \frac{2}{K} \sum_l c_l^2 \varphi''(u_l)^2
		\end{align}
	\end{subequations}
	In the large $K$ limit the variables $\mathcal{O}_\star(\boldsymbol{v})$, $\mathcal{O}(\boldsymbol{x_0},\boldsymbol{x_1},\boldsymbol{x_{01}},\boldsymbol{\eta})$, $\mathcal{O}_1(\boldsymbol{x_0},\boldsymbol{x_1},\boldsymbol{x_{01}},\boldsymbol{v}, \boldsymbol{h})$, $\mathcal{O}_2(\boldsymbol{x_0},\boldsymbol{x_1},\boldsymbol{x_{01}},\boldsymbol{v}, \boldsymbol{h})$ in~\eqref{eq::O} become jointly Gaussian. The mean of all those observables vanishes because of the condition $\sum_l c_l = 0$. 
	The covariance elements uniquely determined by the equilibrium overlaps $Q$, $q_1$, $q_0$, and $r$ are those involving only averages of combinations of $\mathcal{O}$ and $\mathcal{O}\star$
	\begin{subequations}
		\begin{align}
			\Phi_\star &= \langle \mathcal{O}^2_\star \rangle_{v} - \langle \mathcal{O}_\star \rangle_{v}^2 = \mathcal{K}(1,1,1) - \mathcal{K}(1,1,0) \\
			D_0 &= \langle \langle \mathcal{O}\rangle_{x_0, x_1, x_{01}} \mathcal{O}_\star \rangle_{v} - \langle \mathcal{O} \rangle_{} \langle \mathcal{O}_\star \rangle = \mathcal{K}(1,Q,r) - \mathcal{K}(1,Q,0) \\
			\Phi_{Q, q_1} &= \langle \mathcal{O}^2 \rangle_{x_0, x_1, x_{01}, v} - \langle \langle \mathcal{O}\rangle_{x_0}^2 \rangle_{x_1,  x_{01}, v} = \mathcal{K}(Q, Q, Q) - \mathcal{K}(Q, Q, q_1)\\
			\Phi_{q_1, q_0} &= \langle \langle \mathcal{O}\rangle_{x_0}^2 \rangle_{x_1,  x_{01}, v} - \langle \langle \mathcal{O}\rangle_{x_0, x_{01}}^2 \rangle_{x_1, v} = \mathcal{K}(Q, Q, q_1) - \mathcal{K}(Q, Q, q_0)\\
			\Phi_{q_0, 0} &= \langle \langle \mathcal{O} \rangle_{x_0, x_{01}}^2 \rangle_{x_1, v} - \langle \mathcal{O}\rangle
			^2 = \mathcal{K}(Q, Q, q_0) - \mathcal{K}(Q, Q, 0)
		\end{align}
	\end{subequations}
	which match the quantities introduced in Appendix~\ref{sec::Equilibrium} (see e.g.~\eqref{eq::equilibrium_RS_effective_parameters}) via the identification
	\begin{equation}
		\Phi_{a, b} = \Phi(a) - \Phi(b)
	\end{equation}
	The new covariance elements involve the quantities $\mathcal{O}_1$ and $\mathcal{O}_2$ relevant for the determination of the Hessian. Those are
	\begin{subequations}
		\begin{align}
			\Delta_0 &\equiv \langle \mathcal{O}_1^2 \rangle \\
			\Xi_0 &\equiv \langle \mathcal{O}_2^2 \rangle - \langle \langle \mathcal{O}_2\rangle^2_h \rangle_{x_0, x_1, x_{01}, v} \\
			\Sigma_{Q, q_1} &\equiv \langle \langle \mathcal{O}_2\rangle_{h} \mathcal{O} \rangle_{x_0, x_1, x_{01}, v} -  \langle \langle \mathcal{O}_2 \rangle_{h, x_0} \langle \mathcal{O} \rangle_{x_0} \rangle_{x_1, x_{01},v} \\
			\Sigma_{q_1, q_0} &\equiv \langle \langle \mathcal{O}_2 \rangle_{h, x_0} \langle \mathcal{O} \rangle_{x_0} \rangle_{x_1, x_{01},v} - \langle \langle \mathcal{O}_2 \rangle_{h, x_0, x_{01}} \langle \mathcal{O} \rangle_{x_0, x_{01}} \rangle_{x_1, v} \\			
			\Sigma_{q_0, 0} &\equiv \langle \langle \mathcal{O}_2 \rangle_{h, x_0, x_{01}} \langle \mathcal{O} \rangle_{x_0, x_{01}} \rangle_{x_1, v} - \langle \mathcal{O}_2\rangle \langle \mathcal{O} \rangle\\
			D_1 &\equiv \langle \langle \mathcal{O}_2\rangle_{h, x_0, x_1, x_{01}} \mathcal{O}_\star \rangle_{v} - \langle \mathcal{O}_2 \rangle \langle \mathcal{O}_\star \rangle 
		\end{align}
	\end{subequations}
	Those effective order parameters can be all written in terms of the generalized kernel function~\eqref{eq::generalized_nngp_kernel} as
	\begin{subequations}
		\begin{align}
			\Delta_0 &= 
			\mathcal{K}^{(1,1)}(Q,Q,Q) 
			\\
			\Xi_0 &= 2
			\, \mathcal{K}^{(2,2)}(Q,Q,Q) 
			\\
			\Sigma_{Q, q_1} &= 
			\mathcal{K}^{(2,0)}(Q,Q,Q) -\mathcal{K}^{(2,0)}(Q,Q,q_1) 
			\\
			\Sigma_{q_1, q_0} &= 
			\mathcal{K}^{(2,0)}(Q,Q,q_1) - \mathcal{K}^{(2,0)}(Q,Q,q_0)
			\\
			\Sigma_{q_0, 0} &= 
			\mathcal{K}^{(2,0)}(Q,Q,q_0) - \mathcal{K}^{(2,0)}(Q,Q,0)
			\\
			D_1 &= 
			\mathcal{K}^{(0,2)}(1,Q,r)-\mathcal{K}^{(0,2)}(1,Q,0)
	\end{align}
\end{subequations}
Using those covariances one finally finds that the energetic term has the form
\begin{equation}
\mathcal{G}_E = \frac{1}{2} A_2 P^2 + \frac{1}{2} A_1 P - \frac{1}{2} \left\langle \ln \left( 1 + P \Delta_0 \ell''(y, \hat y) \right) \right\rangle_{\beta}
\end{equation}
with the coefficients $A_2$ and $A_1$ being independent of $P$
\begin{subequations}
\begin{align}
	A_2 &=  \frac{\Xi_0}{4} \left\langle \ell'(y, \hat y)^2 \right\rangle_{\beta}   \\
	A_1 &=  - \left\langle \ell'(y, \hat y) \left(\frac{D_1}{\sqrt{\Phi_\star}}v +  \frac{\Sigma_{Q, q_1}}{\sqrt{\Phi_{Q, q_1}}} x_0 + \frac{\Sigma_{q_0, 0} - \frac{D_0 D_1}{\Phi_\star}}{\sqrt{\Phi_{q_0, 0} - \frac{D_0^2}{\Phi_\star}}} x_1 +   \frac{\Sigma_{q_1, q_0}}{\sqrt{\Phi_{q_1, q_0}}} x_{01}\right) \right\rangle_{\!\!\beta} 
\end{align}
\end{subequations}
In the previous expression we have denoted by $\langle \bullet \rangle_\beta$ the energetic equilibrium measure average in the large $K$ limit, which at 1RSB level reads
\begin{subequations}
\begin{align}
	\label{eq::equilibrium_1RSB_measure}
	\left \langle \bullet \right \rangle_{\beta} &\equiv \int Dv Dx_1 \, \frac{\int Dx_{01} \left[\int Dx_0  \, e^{ - \beta \ell(y, \hat y) } \right]^{m} \frac{\int Dx_0  \, e^{ - \beta \ell(y, \hat y) } \bullet}{\int Dx_0  \, e^{ - \beta \ell(y, \hat y) }} }{\int Dx_{01} \left[ \int Dx_0  \, e^{ - \beta \ell(y, \hat y) } \right]^m}  \\
	y &\equiv f_\star\left(\sqrt{\Phi_\star}v\right) \,,\\
	\hat y &\equiv \frac{D_0}{\sqrt{\Phi_\star}} v + \sqrt{\Phi_{Q, q_1}} x_0 + \sqrt{\Phi_{q_0, 0} - \frac{D_0^2}{\Phi_\star}} x_1 + \sqrt{\Phi_{q_1, q_0}} x_{01}\,.
\end{align}
\end{subequations}

\subsection{Square loss case}~\label{sec::Hessian_square_loss}

When one takes the square loss $\ell(y, \hat{y}) = \frac{1}{2}(y - \hat y)^2$ all the integrals in the energetic term are Gaussian and can be solved. The free entropy therefore is given by $\phi = \mathcal{G}_{SI} + \alpha \mathcal{G}_E$ where
\begin{subequations}
\begin{align}
	\mathcal{G}_{SI} &= \frac{1}{2} + \frac{z}{2}P + \frac{1}{2}\ln P \\
	\mathcal{G}_E &= \frac{1}{2}A_2 P^2 + \frac{1}{2}A_1 P - \frac{1}{2} \ln (1 + P\Delta_0)
\end{align}
\end{subequations}
with the coefficients $A_2$ and $A_1$ that read
\begin{subequations}
\begin{align}
	A_2 &= \frac{\Xi_0}{4(1+\beta \Phi_{Q, q_1})^2} \left[ \Phi_{Q, q_1} (1+\beta \Phi_{Q, q_1})	+\frac{\Phi_{q_1, q_0}}{B} + \frac{C}{B^2} \right] \\
	A_1 &= -\frac{1}{(1+\beta \Phi_{Q, q_1})^2}\left[ (1+\beta \Phi_{Q, q_1}) \left( \Sigma_{Q, q_1} + \frac{T}{B} \right)
	-\frac{\beta\Sigma_{Q, q_1}}{B}\left(\Phi_{q_1, q_0}+\frac{C}{B}\right)
	-\Sigma_{q_1, q_0}\,\frac{m\beta}{B^2}\,C	\right] \\
	B &= 1+\frac{m\beta \Phi_{q_1, q_0}}{1+\beta \Phi_{Q, q_1}}
\end{align}
\end{subequations}
The constants $C$ and $T$ depend on the task, i.e. they differ from the classification ($f_\star(\hat y) = \mathrm{sign}(\hat y)$) to the regression case ($f_\star(\hat y) = \hat y$).
In classification one has
\begin{subequations}
\label{eq::CT_classification}
\begin{align}
	C_\mathrm{class} &= 1 + \Phi_{q_0, 0}-2 D_0  \sqrt{\frac{2}{\pi \Phi_\star}} \\
	T_\mathrm{class} &= \Sigma_{q_0, 0}+\Sigma_{q_1, q_0}- D_1 \sqrt{\frac{2}{ \pi \Phi_\star}}
\end{align}
\end{subequations}
whereas in regression
\begin{subequations}
\label{eq::CT_regression}
\begin{align}
	C_\mathrm{reg} &= \Phi_\star+\Phi_{q_0, 0}-2D_0 \\
	T_\mathrm{reg} &= \Sigma_{q_0, 0}+\Sigma_{q_1, q_0} - D_1
\end{align}
\end{subequations}
The saddle point equation for $P$ in both cases is therefore a cubic equation
\begin{equation}
\label{eq::cubic_equation}
\begin{split}
	1+ \tilde{a}_1(z) P + \tilde{a}_2(z) P^2 + \tilde{a}_3 P^3 = 0
\end{split}
\end{equation}
with coefficients given by
\begin{subequations}
\begin{align}
	\tilde{a}_1(z) &= z + (1 - \alpha) \Delta_0 + \alpha A_1\\
	\tilde{a}_2(z) &= z \Delta_0 + \alpha(2A_2 + A_1 \Delta_0 ) \\
	\tilde{a}_3 &= 2 \alpha \Delta_0 A_2 \,.
\end{align}
\end{subequations}
Note that the coefficients in the RS case can be obtained by setting $q_1 = q_0 = q$ that is $\Phi_{q_1, q_0} = \Sigma_{q_1, q_0} = 0$. 

Equation~\eqref{eq::cubic_equation} can be therefore solved exactly. Indeed the cubic equation has either three real roots, corresponding to a vanishing spectral density, or one real root and a complex-conjugate pair. In the latter case, because of~\eqref{eq::spectrum_P} the physical solution is the root with negative imaginary part, so that $\rho_{\alpha \mathcal H}(z)=-(1/\pi)\operatorname{Im}P(z)\ge 0$. 
Finally, recalling that the computation above concerns $\alpha \mathcal H$, we have $\mathcal{R}_{\alpha \mathcal H}(z)=-P(z)$ and $\mathcal R_{\mathcal H}(z)=\alpha \mathcal R_{\alpha \mathcal{H}}(\alpha z)$. Therefore~\eqref{eq::cubic_equation} evaluated at $\alpha z$, yields equation~\eqref{eq:cubic_main} of the main text with the identification
\begin{equation}
a_1(z)=-\frac{\widetilde a_1(\alpha z)}{\alpha},\qquad
a_2(z)=\frac{\widetilde a_2(\alpha z)}{\alpha^2},\qquad
a_3=-\frac{\widetilde a_3}{\alpha^3}.
\end{equation}

\subsection{Zero temperature limit}

Remind that in the large $\beta$ limit we have to impose the following scalings
\begin{subequations}
\begin{align}
	q_1 &= Q - \frac{\delta q}{\beta} \\
	m &= \frac{\delta m}{\beta}
\end{align}
\end{subequations}
This implies the following scalings in $\beta$ for the effective order parameters
\begin{subequations}
\label{eq::effop_scalings}
\begin{align}
	\Phi_{Q, q_1} &= 
	\frac{\delta \Phi}{\beta} \\
	\Sigma_{Q, q_1} &= \frac{\delta \Sigma}{\beta}
\end{align}
\end{subequations}
It is easy to show by Taylor expansion that $\delta \Phi$ and $\delta \Sigma$ are expressed in terms of the generalized kernel function in~\eqref{eq::generalized_nngp_kernel} as
\begin{subequations}
\begin{align}
	\delta \Phi   &= \mathcal{K}^{(1,1)}(Q,Q,Q) \, \delta q\\
	\delta \Sigma &= \mathcal{K}^{(3,1)}(Q,Q,Q) \, \delta q 
\end{align}
\end{subequations}
Note that the other effective order parameters stay finite in the large $\beta$ limit. One finally finds
\begin{equation}
	\begin{split}
		\lim\limits_{\beta \to \infty} \mathcal{G}_E &= - \frac{P}{2} \left\langle \ell'(y, \delta \hat y_\star) \left(\frac{D_1}{\sqrt{\Phi_\star}}v +  \frac{\delta \Sigma}{\sqrt{\delta \Phi}} x_0^\star(v, x_{01}, x_1) + \frac{\Sigma_{q_0, 0} - \frac{D_0 D_1}{\Phi_\star}}{\sqrt{\Phi_{q_0, 0} - \frac{D_0^2}{\Phi_\star}}} x_1 +  \frac{\Sigma_{Q, q_0}}{\sqrt{\Phi_{Q, q_0}}} x_{01}\right) \right\rangle_{\!\!\infty} \\
		&+ \frac{\Xi_0 P^2}{8} \left\langle \ell'(y, \delta \hat y_\star)^2 \right\rangle_{\infty} - \frac{1}{2} \left\langle \ln \left( 1 + P \Delta_0 \ell''(y, \delta \hat y_\star) \right) \right\rangle_{\infty}
	\end{split}
\end{equation}
where
\begin{subequations}
\begin{align}
	x_0^\star(v, x_{01}, x_1) &\equiv \argmax_{x_0} \left[ - \frac{x_0^2}{2} - \ell \left(y(v), \delta y(v,  x_0, x_{01}, x_1) \right) \right] \\
	\delta \hat y(v, x_0, x_1, x_{01}) &\equiv \frac{D_0}{\sqrt{\Phi_\star}} v + \sqrt{\delta \Phi} x_0 + \sqrt{\Phi_{q_0, 0} - \frac{D_0^2}{\Phi_\star}} x_1 + \sqrt{\Phi_{Q, q_0}} x_{01} \\
	y(v) &\equiv f_\star\left(\sqrt{\Phi_\star}v\right) \\
	\delta \hat y_\star &= \delta \hat y(v, x_0^\star(v, x_{01}, x_1),  x_1, x_{01})
\end{align}
\end{subequations}
with the equilibrium measure in the large $\beta$ limit being
\begin{equation}
\label{eq::equilibrium_1RSB_measure_largebeta}
\left \langle \bullet \right \rangle_{\infty} \equiv \int Dv Dx_1 \, \frac{\int Dx_{01}  \, e^{ - \delta m \, \ell(y, \delta \hat y_\star) } \, \bullet }{\int Dx_{01} \, e^{ - \delta m \,  \ell(y, \delta \hat y_\star) }} 
\end{equation}
The replica symmetric case can be again recovered setting $\Phi_{Q, q_0} = \Sigma_{Q, q_0} = 0$.

\subsubsection{Square loss case}

In the square loss case the limit can be worked out directly from the expressions given in section~\ref{sec::Hessian_square_loss}. What will change is the expression of the coefficients $A_1$ and $A_2$. They are given by
\begin{subequations}
\begin{align}
	A_2 &= \frac{\Xi_0}{4(1+ \delta \Phi)^2} \left[ \frac{\Phi_{Q, q_0}}{B} + \frac{C}{B^2} \right] \\
	A_1 &= -\frac{1}{(1+ \delta \Phi)^2}\left[ (1+\delta \Phi) \frac{T}{B} 
	-\frac{\delta\Sigma}{B}\left(\Phi_{Q, q_0}+\frac{C}{B}\right)
	-\Sigma_{Q, q_0}\,\frac{\delta m}{B^2}\,C	\right] \\
	B &= 1+\frac{\delta m\Phi_{Q, q_0}}{1+ \delta \Phi}
\end{align}
\end{subequations}
where we remind that the constants $C$ and $T$ depend on the task and are given by~\eqref{eq::CT_classification} and~\eqref{eq::CT_regression} respectively in classification and regression, and tend to a finite limit for large $\beta$.

\subsection{Left and right edges of the spectrum of the Hessian}

We remind the reader that the Stieltjes transform of the Hessian and the spectrum are respectively given by  (see Appendix~\ref{app::Gibbs_Hessian})
\begin{subequations}
\begin{align}
	\mathcal{R}_{\alpha \mathcal{H}}(z) &= -2 \frac{\partial}{\partial z} \phi(z) = - P \\
	\rho_\mathcal{\alpha H}(z) &= - \frac{1}{\pi} \text{Im} P \,,
\end{align}
\end{subequations}
and the value of $P$ is found maximizing $\phi(z)$ with respect to $P$, which gives
\begin{equation}
z = -\frac{1}{P} + \alpha \left[-2A_2 P  - A_1 +\left\langle \frac{\Delta_0 \ell''(y,\hat{y})}{1+P\Delta_0 \ell''(y,\hat{y})} \right\rangle_\beta \right]
\end{equation}
The edges of the spectrum can be obtained by looking for the points where the derivative of $z(P)$ is equal to zero, as beyond these points $z(P)$ is not defined for real $P$.
Setting $\frac{dz}{dP}=0$ we get the two solutions (corresponding to the left and right edges)
\begin{equation}
P_{\pm} = \pm \frac{1}{\sqrt{\alpha}} \Biggl[ 2A_2 + \left\langle \left(\frac{\Delta_0 \ell''(y,\hat{y})}{1+P_\pm \Delta_0 \ell''(y,\hat{y})} \right)^2\right\rangle_\beta \Biggr]^{-1/2} \,.
\end{equation}

\subsection{A comment on the validity of the Annealed Ansatz}

It is possible to generalize the previous computation which was valid for an annealed structure of the order parameters~\eqref{eq::annealed_ansatz}, to a replica symmetric one
\begin{subequations}
\label{eq::RS_ansatz}
\begin{align}
	p_l^{cd} &= P \delta_{cd} + (1-\delta_{cd}) p \\
	\hat p_l^{cd} &= - \hat P \delta_{cd}  + (1-\delta_{cd}) \hat p\\
	t_l^{ac} &= T \delta_{a1} + (1-\delta_{a1})t \\
	\hat t_l^{ac} &= \hat T \delta_{a1} + (1-\delta_{a1}) \hat t \\
	o_l^{c} &= o \\	
	\hat o_l^{c} &= \hat o
\end{align}
\end{subequations}
The computation is however long and tedious (especially when dealing with the large width limit) and we do not report it here. However one can analyze straightforwardly the saddle point equations. First the equation for the conjugated parameters $\hat P$, $\hat T$, $\hat t$ and $\hat o$ can be solved explicitly in terms of $P$, $p$, $T$, $t$, $o$. One finds then that the free entropy function is a quadratic form of the variables $T$, $t$ and $o$. The only solution of the saddle point equations is therefore $T=t=o=0$. Inserting this solution inside the free entropy one also finds that $p=0$ is always a solution. In this way one finds that 
\begin{equation}
t=\hat t = T = \hat T = p = \hat p = 0 \,.
\end{equation}
This tells us that the annealed ansatz is the correct one.

\section{Details of the numerical experiments}
\label{app:numerics}

We validate the theory with full-batch gradient-descent simulations of the model
of Section~\ref{sec:model}, using $N=2000$, $K=20$, and $\varphi=\operatorname{erf}$.
For each $\alpha$ we draw a teacher 20 times, generate $P=\lfloor\alpha N\rfloor$ i.i.d.\ standard Gaussian training inputs and an independent test set of size 10000, and minimize the regularized objective~\eqref{eq:loss_main} with nominal weight decay $\lambda$ and learning rate $\eta_0\in[0.5,1.0]$. Training runs for up to $E_{\max}\in[10^4,10^5]$ epochs, with early stopping when the relative change of the loss and of the order parameters $Q=\lVert\w\rVert^2/N$ and $r=\w\cdot\w^\star/N$ all fall below $\varepsilon=10^{-8}$ between two consecutive checkpoints (checked every $100$ epochs). At the end of training we evaluate the Hessian of the unregularized loss in closed form, exploiting the committee structure, and diagonalize the full $N\times N$ matrix exactly, without any low-rank or stochastic approximation. The experiments are implemented in Julia,
using Flux.jl for training, ForwardDiff.jl for $\varphi',\varphi''$, and
LinearAlgebra.jl for the diagonalization.

\end{document}